\documentclass{article}
\usepackage{style} 

\title{\Large Social Networks and Internal Migration: \\ Evidence from Facebook in India\thanks{\footnotesize We are grateful to Michael Greenstone, Michael Kremer, Michael Dinerstein, numerous seminar participants, and especially Jordan Rosenthal-Kay for critical guidance. This research was facilitated through a
research consulting agreement between the academic author (Sahai) and Meta. Bailey is
an employee at Meta. All errors are our own.}}
\author{
\begin{tabular}[t]{c@{\extracolsep{1em}}c} 
 Harshil Sahai\thanks{Corresponding author. Kenneth C. Griffin Department of Economics, The University of Chicago (\url{harshil@uchicago.edu}).} & Michael Bailey\thanks{Meta Platforms, Inc.}
\end{tabular}
}
\date{December 2023 \\ \href{https://www.harshil.com/s/sahai_india_migration.pdf}{Latest Version}}
\RequirePackage{etex} 

\begin{document}

\defcitealias{who2018world}{WHO}
\maketitle

\begin{abstract}
\noindent Despite potentially large economic returns, rates of internal migration remain low in many developing countries. This paper uses new, de-identified data from Facebook to quantify the role of social networks in explaining this development puzzle. We study this question in India, a country that exhibits substantial wage dispersion across regions but remains relatively under-urbanized. Detailed records of nearly 20 million individuals on the evolution of social connections and residential choice reveal that networks and migration are strongly linked. Across several identification strategies, a model of migration suggests that social networks account for roughly 20\% of the relationship between migration and distance. We develop a simple, static model of spatial equilibrium, which suggests that equalizing social connections across locations increases average wages by 3\% (24\% for the bottom wage-quartile) through increased migration. This impact is larger than fully removing the marginal effect of distance in migration decisions, akin to building rapid transport infrastructure. Taken together, our data suggest that -- by reducing migration frictions -- increasing social connections across space may have considerable economic gains. We provide suggestive evidence for economic and emotional support mechanisms underlying network effects and show that college attendance can boost the size and diversity of social networks by 20\%.
\end{abstract}

\clearpage



\onehalfspacing 

\section{Introduction}

Gravity models of migration continue to document large moving costs that increase with distance -- reductions in such frictions may have substantial impacts to aggregate productivity and welfare.\footnote{See, for example, \cite{bcm14}, \cite{bm19}, \cite{dnr18}, and \cite{tz19}.} However, these barriers to migration can be explained by a multitude of forces, such as physical moving costs, culture and language differences, skill transferability, limited information, or social networks.\footnote{See, for example, \cite{anderson11}, \cite{mr16}, and \cite{schwartz73}.} In order to address this with policy, understanding the source of these moving costs is therefore of central importance. While theory and evidence suggests social networks are particularly important determinants of residential choice, little work has quantified what portion of moving costs are explained by networks, especially so in low-income settings where data is relatively scarce and informal networks may be relatively more important.\footnote{In the sociology literature see \cite{b83}, \cite{b89}, and \cite{mm64}. In economics, see \cite{munshi20} for a review.} 

In this paper, we provide new evidence on this question from the world's largest online social network, Facebook, in its largest consumer market, India. Using aggregated and de-identified data on nearly 20 million individuals, we develop and estimate a framework to quantify how social networks influence migration decisions, and what this may mean for aggregate productivity.


We start by presenting new data on nearly 20 million individuals that record yearly residential decisions and social networks over four years.\footnote{Meta, Facebook’s parent company, has a home prediction model that determines home regions based on self-reported hometowns and device and connection information. Similar profile information has been used to estimate migration in \cite{hsaw16} and \cite{csba20}.} This reveals two empirical patterns: first, social capital -- as defined by the number of Facebook Friends and the average income of where they live -- is concentrated among the rich, the educated, and those who migrate; and second, both migration patterns and social connections are concentrated toward richer areas. These facts suggest that social networks and internal migration are strongly linked.

To quantify this relationship, we then develop a gravity model of migration that includes preferences for city-specific social networks, which is estimated using microdata on millions of yearly residential decisions in India. We develop three different identification strategies to address endogeneity in social networks. First, we exploit the fine granularity of data and estimate the model using highly restrictive fixed effects, up to the origin-by-destination-by-year level (\cite{bct19}). Second, we develop a novel instrument that exploits productivity shocks to locations combined with individual network structure to deliver plausibly random variation in social connections across locations. Third, we administer a hypothetical survey of 50,000 Facebook users to capture preferences for cities that hold both wages and networks fixed across space (``amenities''). Across all three approaches, we find consistent evidence of large network effects: individuals are indifferent between a 10\% increase in average destination wages and a 12-16\% increase in destination social networks. In our preferred specification, accounting for networks in the model reduces the migration-distance relationship by 19\%, a substantial share of total moving costs. Further, the model suggests there are strong interactions between network effects and distance. Increasing social connections to a destination from 0 to 5 reduces the marginal cost of distance by 40\%. This suggests that social networks act to effectively lower moving costs.

Finally, we integrate our model of migration into a simple, static spatial equilibrium model of local labor markets, which incorporates both agglomeration and congestion forces (\cite{rr17}). The model includes endogenous wages and amenities that adjust to ensure equilibrium across labor supply and demand. We use this framework to test the economic impacts of various counterfactuals. We find that, in equilibrium, removing the marginal cost of distance in the migration decision increases average wages by 3\% (17\% for the bottom wage-quartile) through increased migration. This is akin to an experiment that builds rapid transport infrastructure, dramatically reducing moving costs that are related to distance. We find that even larger gains can be achieved by equalizing or reallocating social networks across cities, which deliver a 3-7\% increase in average wages (24-28\% for the bottom wage-quartile). This suggests there are potentially large and progressive economic gains from increasing social connections across space, through the reduction in effective moving costs.

What drives such strong preferences for social networks? Networks may affect migration through multiple channels (\cite{b89}). For example, informal networks provide economic support that may correct for market imperfections commonly seen in developing countries (e.g. credit, insurance, information, housing). Alternatively, social connections may provide emotional support that simply reflect individual taste (e.g. leisure, socialization). We find suggestive evidence for both mechanisms. Network effect heterogeneity suggests that economic improvements in the destination -- formal banking expansion, higher wages, or migrant-friendly policy -- deliver marked reductions in the migration-network elasticity. This points to an economic support mechanism in driving part of the relationship between migration and networks. Policy that targets incomplete credit or insurance markets may thus have additional benefits through a reduced reliance on informal networks in determining migration.

We test for other mechanisms by administering qualitative surveys on social support among 50,000 Facebook users. We find that, as their most important source of support, individuals are 20\% more likely to list emotional support (such as happiness from spending time together or talking during stressful times) than economic support (such as information about job opportunities or access to informal loans). This suggests that economic policy may not fully eliminate preferences for social networks.

While the \emph{effect} of networks may persist, social networks themselves can grow. We examine the role of education in expanding social capital using data on the evolution of networks over the life cycle. By comparing attendee versus non-attendees, difference-in-difference estimates suggest college attendance delivers a 20\% increase in the number of social connections and the number of unique cities to which individuals are connected. This suggests education may have additional economic returns through its impact on social capital, and thus the reduction in effective moving costs.

This paper contributes to several strands of the development and labor economics literature. There is a rich theoretical and empirical literature which studies the potential role of social networks in influencing migration decisions. For example, prior work has studied how immigrant communities influence immigration patterns in the US (\cite{beaman11}, \cite{munshi03}); and how caste networks influence village out-migration in India (\cite{mr16}). However, due to extremely limited data linking migration decisions with social networks, little work has directly quantified the magnitude of these network effects. A recent exception is \cite{bct19}, which uses cell phone location and call data in Rwanda to estimate the relationship between region-specific monthly migration and phone contacts. 

We add to this literature in several ways. First, to our knowledge, we estimate effects on the largest, most granular sample ever compiled on networks and migration, across nearly 20 million individuals and more than 5,000 cities. Second, we improve identification by estimating effects across three approaches, including restrictive fixed effects, instruments that exploit local productivity shocks combined with individual network structure, and hypothetical survey questions that separately estimate amenities. Third, we incorporate the migration model into a spatial equilibrium framework that allows us to study the aggregate productivity implications of network-related counterfactuals.

There is a separate literature on the spatial misallocation of labor and the economic implications of reducing moving costs. For example, \cite{bm19} uses microdata in Indonesia to measure moving costs in a gravity model with costly migration, and assesses the aggregate productivity implications of reducing these costs in static spatial equilibrium. \cite{dnr18} shows that in a dynamic spatial framework with costly trade and migration, fully removing moving costs would increase welfare threefold. While these moving costs may be large and economically meaningful, what is less known is their composition. Understanding the source of these moving costs has clear policy relevance. For example, \cite{mo18} finds that improvements in transport infrastructure in Brazil have only modest impacts to aggregate productivity via migration. To the best of our knowledge, no prior work has estimated models of spatial equilibrium that account for social networks, or quantified the share of moving costs explained by networks. We therefore contribute by providing the first such study, and showing that roughly one-fifth of the gravity relationship in India can be explained by social networks.

This paper proceeds as follows. Section \ref{sec:data} describes our data; Section \ref{sec:desc} presents descriptive findings of migration and social networks in India; Section \ref{sec:model} develops a model of location choice with social networks and discusses identification; Section \ref{sec:modelresults} reports model estimates on the effects of social networks; Section \ref{sec:ge} presents a simple model of spatial equilibrium and simulates various policy counterfactuals; Section \ref{sec:mech} provides suggestive evidence on mechanisms; and Section \ref{sec:discussion} concludes.

\section{Data description}
\label{sec:data}

This paper uses three main sources of data: (i) aggregated and de-identified data on individual social networks and home city predictions in India from Facebook.com, the world's largest online social network; (ii) administrative data from the Government of India on local labor markets; and (iii) supplemental data from qualitative surveys among Facebook users in India.

\subsection{Facebook data in India}
Facebook.com is the world's largest online social network, with nearly 3 billion monthly active users around the world (\cite{meta22}). The website allows users to connect and communicate with friends and family. India is Facebook's largest consumer market, with nearly 400 million users, roughly 30\% of the population (\cite{statista21}). When restricting to young, working-age men in India (between 18 and 34 years old), more than 90\% of these individuals have a Facebook account.\footnote{These are estimated based off author calculations using age and gender distributions of Facebook users from \cite{statista20} and age-specific India population estimates from the World Bank in 2020.} In this paper, we use data from Facebook to construct an individual panel across nearly 20 million individuals over 4 years of (i) the complete social graph across individuals and (ii) annual migration decisions across cities for each individual. 

On the platform, users can request Facebook connections (or ``Friends'') from any other individual on the platform at any time. By accepting these requests (or others accepting a sent request), users social networks may grow over time. These connections capture a large set of potential social networks: friends, family, coworkers, acquaintances, etc. Because connections require consent of both individuals, they are primarily between individuals who interact in person and therefore resemble real-world social networks (\cite{j13}). We observe pairwise connections between all users at the end of each year. Similar Facebook data has been used to study the role of social networks in housing markets and international trade, among other topics (\cite{bcksw18}; \cite{bcks18}).

Meta, Facebook’s parent company, has a home prediction model that determines users' city of residence based on device and connection information in addition to self-reported data. Similar profile information has been used to estimate migration in \cite{hsaw16} and \cite{csba20}. This allows us to observe individual migration patterns to different cities over time (``home city predictions'') as well as origin locations from which initial migration would take place (self-reported ``hometowns'').\footnote{Users may also self-report age, gender, and education history (e.g. high school, college, graduate school). We also observe what device model is used to access the platform. This device data has been shown in prior research to correlate strongly with wealth (\cite{cfcb21}). We exploit these individual characteristics in our descriptive analysis and sample restrictions.} 

We restrict our data to a subset of all Facebook users due to data limitations as well as relevance to our focus on labor migration. We first restrict to working age men in India, between 18 and 65 years old, the demographic that dominates labor migration in India. We then limit our analysis to users who satisfy the following criteria: (i) have a valid city prediction for all years between 2014 and 2017 and (ii) live in their self-reported hometown in 2014, the initial period.\footnote{To construct annual city predictions, we collect home city predictions for each user for each day from 2014 to 2017. We collapse this to an annual panel by choosing the modal city across all days in each year for which the user was active on Facebook within a 7 day interval. We restrict to a 7-day active window to ensure consistent city predictions.} We construct pair-wise connections between any two users in this sample at the end of each year. This ensures that growth in social networks is not due to entry and exit of users on Facebook, but changes in ties between existing users only. This results in a consistent panel of 17 million individuals over 4 years from 2014-2017, with city locations and pairwise social networks across nearly 5,000 cities. To the best of our knowledge, this is the largest, most granular data ever compiled on both social networks and migration in India.

\paragraph{Representativeness}

We benchmark our constructed sample from Facebook against available statistics from Indian administrative data on local population and aggregate migration patterns. First, we compare the population distribution across Indian districts to home city predictions in our Facebook sample. Second, we compare the distribution of average wages across Indian districts with the average manufacturer price of the phone device used by individuals in our Facebook sample. Third, we compare net migration rates across Indian states against those computed using changes in city locations in our Facebook sample.

\begin{figure}[H]
\caption{Benchmarking Facebook Data in India}
\label{fig: benchmark}
\centering
\begin{subfigure}{0.55\textwidth}
    \caption{Population by district}
    \label{fig: pop}
    \includegraphics[width=\textwidth]{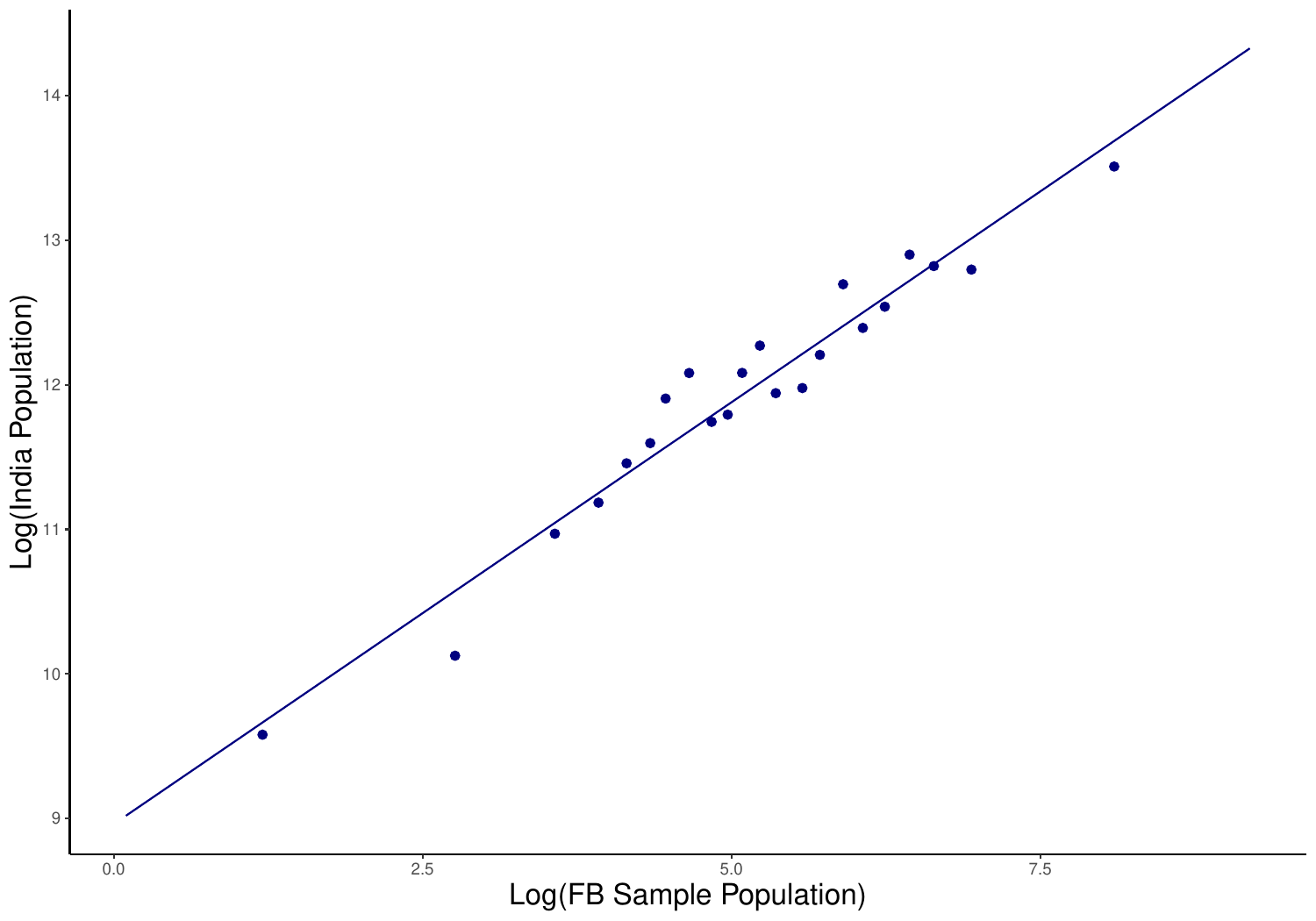}
\end{subfigure}
\hfill
\begin{subfigure}{0.4\textwidth}
    \caption{Wages and device prices}
    \label{fig: w_price}
    \includegraphics[width=\textwidth]{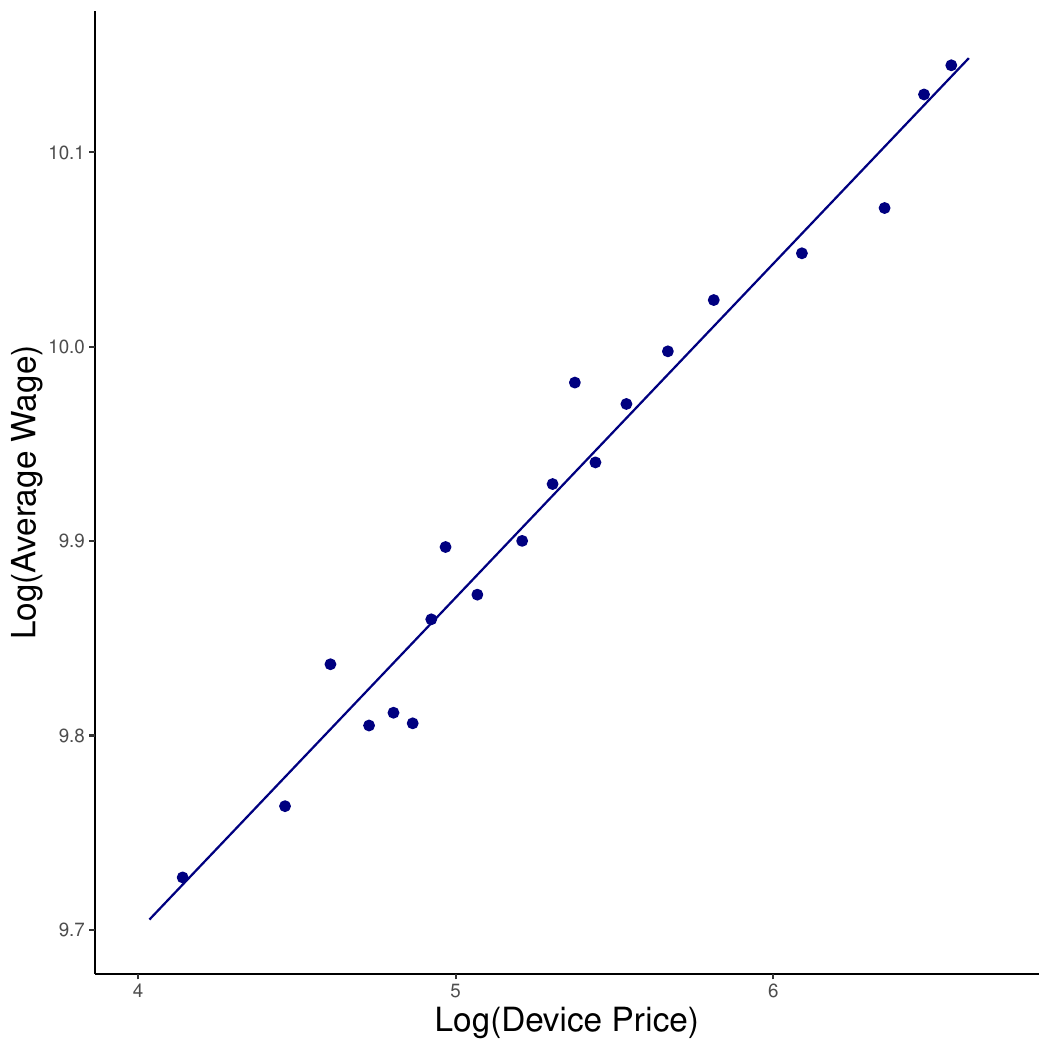}
\end{subfigure}
\hfill
\begin{subfigure}{0.7\textwidth}
    \caption{Net migration rates}
    \label{fig: net}
    \includegraphics[width=\textwidth]{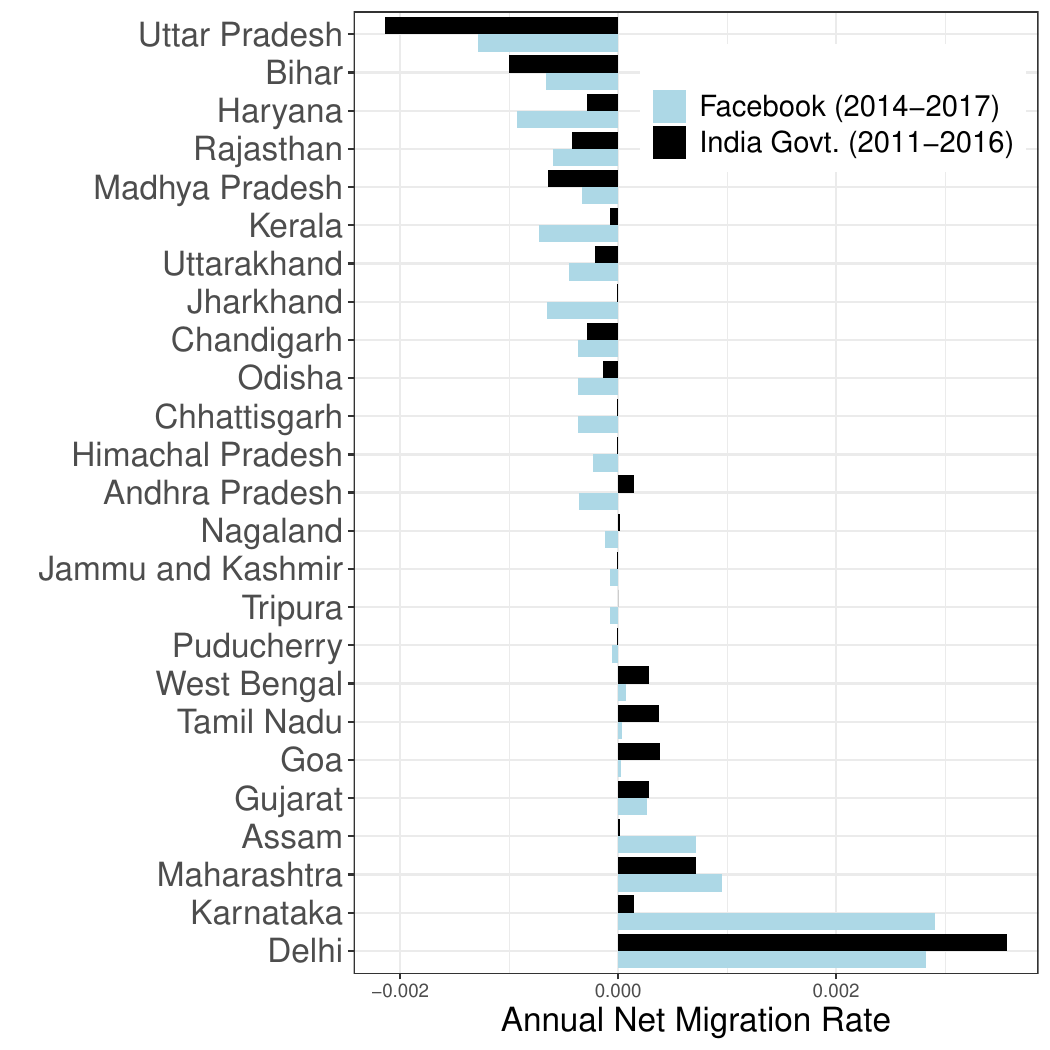}
    \vspace{0.5em}
\end{subfigure}
\begin{notes}
This figure benchmarks data from our Facebook sample against Indian government statistics. Panel A plots a binscatter across Indian districts of the (log) population according to the 2011 Indian census (y-axis) and the population in our sample (x-axis). Panel A plots a binscatter across Indian districts of the (log) average wage according to the 2017-18 Periodic Labour Force Survey by the Government of India (y-axis) and the Average Selling Price (ASP) of the device used by individuals in our sample who reside in that district (x-axis). Panel C reports net in-migration rates across Indian states (number of individuals moving into the state from another state over the total population) from the Government of India railway data between 2011 and 2016 (black) and those computed from our Facebook sample between 2014 and 2017 (blue).
\end{notes}
\end{figure}

Figure \ref{fig: benchmark} compares data from our Facebook sample against administrative statistics. Panel A shows that population across more than 600 districts in India is highly correlated with the population of our Facebook sample using city predictions $(\rho=0.8)$. Panel B shows that the average wage by district is correlated with the Average Selling Price (ASP) of devices used by individuals in our sample ($\rho=0.4)$. Lastly, Panel C reports net in-migration rates for each state in India, comparing 5-year rates from Indian government data (black) to 4-year rates from our Facebook sample (blue). The Facebook data replicates large in-flow states like Delhi as well as large out-flow states like Uttar Pradesh. While differences exist (potentially reflecting differences in sample or time span), the correlation is strong $(\rho=0.8)$. We conclude that our Facebook sample accurately captures aggregate, long-term migration patterns in India.

Table \ref{tab: summary} reports summary statistics for our Facebook sample in comparison to available data in India. The average individual in our Facebook sample is young, roughly 28 years old, and lives in a district with average wages of \$9,300 per year, though considerable variation exists. On average, individual social networks span over 100 friends and 25 cities. Each year, roughly 11\% of individuals move out of their residence city, 8\% out of their district, and 2\% out of their state. Conditional on migration, the average migrant moves roughly 230 kilometers away from their previous residence, to a city with an average wage that is roughly \$2,200 higher. Migrants move with 85\% probability to destinations where they have at least 1 Facebook friend. On average, migrants choose destinations with nearly 23 friends, roughly 20\% of individuals' entire social network.

While our data captures a large section of India's population, this sample is selected in two ways. First, we restrict to a consistent sample of users for which we observe locations and social networks over 4 years. Second, although Facebook is free to use, internet access, device ownership, and personal preferences limit the reach of online social networks like Facebook. On net, we expect that relatively poorer individuals would be underrepresented in our data. Indeed, we find that, compared to the Indian average, the average Facebook user in our sample lives in an area with a 41\% higher mean wage, owns a mobile phone with a 38\% higher price, and has twice the probability of migrating out of their district or state. Importantly, while it may not represent the average Indian, our data span across each of 600+ administrative districts in India, and are more representative of the roughly 30\% of Indians who are on Facebook.


\begin{table}[H] \centering 
  \caption{Summary Statistics: Facebook Sample vs. India} 
  \vspace{-1em}
  \label{tab: summary} 
\begin{tabular}{@{\extracolsep{5pt}}lcccc} 
\\[-1.8ex]\hline 
\hline \\[-1.8ex] 
& \multicolumn{2}{c}{Facebook Sample} & \multicolumn{2}{c}{India} \\
 & \multicolumn{1}{c}{Mean} & \multicolumn{1}{c}{St. Dev.}  & \multicolumn{1}{c}{Mean} & \multicolumn{1}{c}{St. Dev.}\\ 
\hline \\[-1.8ex] 
\emph{Demographics} & & & & \\
\quad Age & 27.8 & 8.1 & 36.2 & 10.9 \\ 
\quad Avg. Wage of Residence (USD) & 9,281.2 & 4,235.8 & 6,579.4 & 3,807.2 \\ 
\quad Device Price (USD)  &256.4  & 186.4 & 153 & \\ 
\emph{Networks} & & & & \\
\quad Network Size  & 105.0 & 156.2 & & \\ 
\quad No. Cities & 25.1 & 23.8 & &\\ 
\emph{One Year Migration}  & & & &\\
\quad Migrated City &  0.11 & 0.31 & &\\ 
\quad Migrated District & 0.08 & 0.27 & 0.04 & \\ 
\quad Migrated State $>$200KM & 0.02 & 0.13 & 0.01 &  \\ 
\emph{If Migrated}  & & & &\\
\quad Distance Migrated (KM) &  227.4 & 359.3 & &\\ 
\quad Annual Wage Gain (USD) & 2,204.2 & 4,758.6 & & \\ 
\quad Destination Has Friend & 0.85 & 0.36 & &\\ 
\quad Destination Network Size & 22.5 & 53.3 & &\\ 
\hline \\[-1.8ex] 
Observations & \multicolumn{2}{c}{17M} & \multicolumn{2}{c}{--} \\
\hline \\[-1.8ex] 
\end{tabular}
\begin{notes}
This table reports means and standard deviations for individual variables across our Facebook sample (left) and available Indian government statistics (right). Facebook statistics come from author calculations for individuals in our Facebook sample in 2017. India statistics on age and residence wage comes from the 2017-18 round of the Periodic Labour Force Survey by the Government of India, restricted to males between 18 and 65 years old to align with the Facebook sample. Indian statistics on device price and migration come from estimates from a 2017 Nielson survey and the 2011 Census of India, respectively.
\end{notes}
\end{table} 


\subsection{Indian administrative data}
While data from Facebook captures migration and social networks, we require data on local labor markets to study wage implications. To this end, we use additional microdata from the 2017-18 round of the Periodic Labour Force Survey (PLFS), administered by the Government of India. This large-scale government data collection effort captures a representative sample of nearly half a million people in the country, including outcomes such as employment status, industry, occupation, and wages.

India is composed of roughly 600 ``districts'' spread across 27 states. The PLFS can be aggregated to the district-level to capture spatial wage heterogeneity, and ultimately, to what extent wages drive migration patterns in India. To the best of our knowledge, this is the finest spatial granularity for which wage data exists in India.

\begin{figure}[H]
\caption{Wages and Urbanization Across India}
\label{fig: wage_het}
\centering
\includegraphics[width=\textwidth]{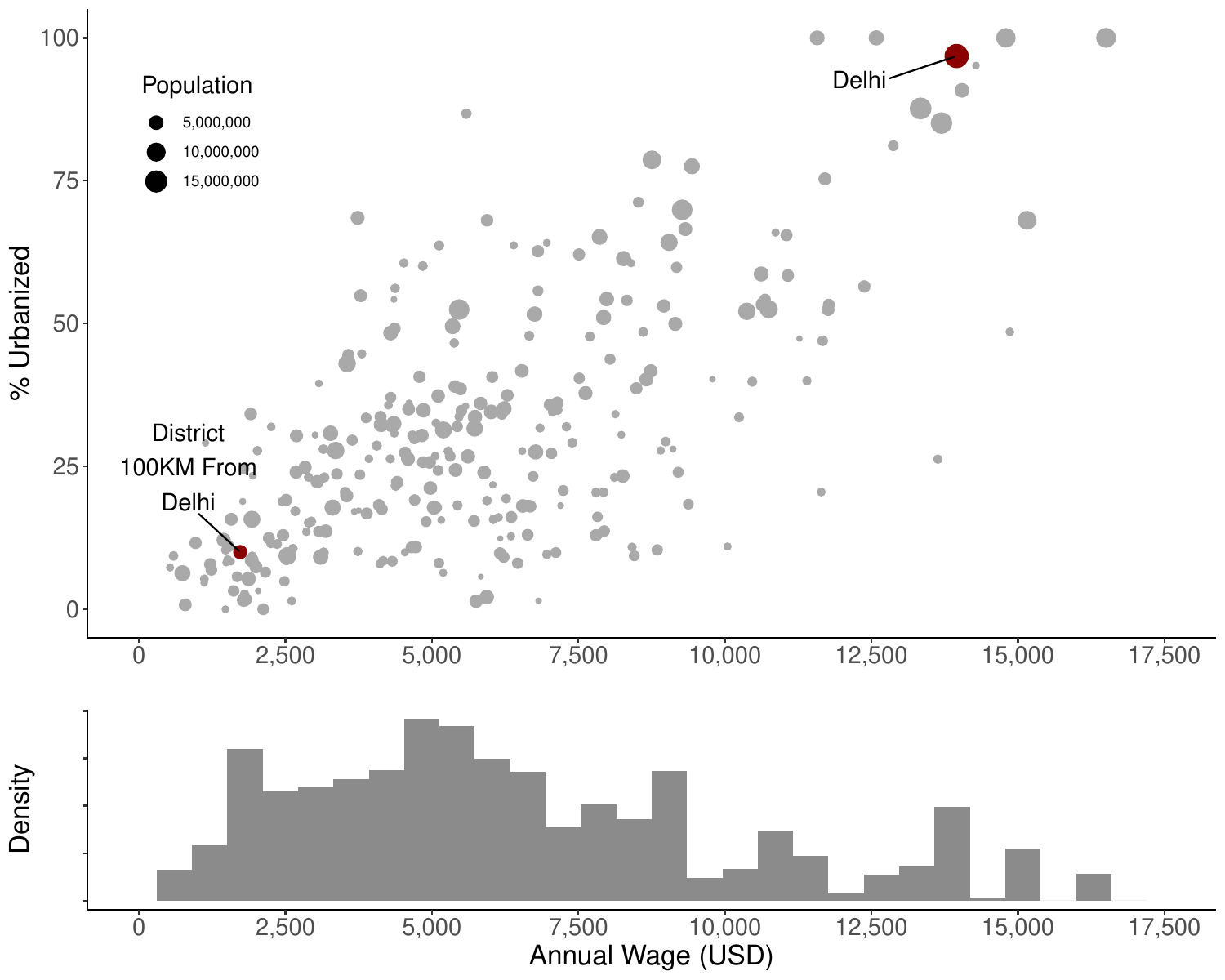}
\begin{notes}
The top panel of this figure presents a scatter-plot at the district-level of average annual wages (x-axis) and fraction urbanized (y-axis); the size of each bubble is proportional to the population of the district. The bottom panel shows the population distribution across average wages. Data come from the 2017-18 round of the Periodic Labour Force Survey (PLFS) by the Government of India, averaged to the district-level using survey weights. ``\% urbanized'' is defined as the fraction of the district population living in a town with population density exceeding 10,000 persons per square kilometer, as in the 2011 Census of India. The distribution of average wages is computed using (survey-weighted) counts from the PLFS.
\end{notes}
\end{figure}

In Figure \ref{fig: wage_het}, we plot district-level average wages against urbanization rates. We see a strong relationship between urbanization and wages, highlighting the degree of wage dispersion in India, with more than a 5x wage premium of being fully urban versus fully rural. For example, the urbanized, capital city of Delhi boasts an average wage of nearly \$15,000 per year, while a rural district 100 kilometers away earns less than \$2,500 per year on average. Despite these differences, the population remains concentrated among the lower-wage and lower-urbanized districts. Indeed, compared to other less-developed countries, India has one of the lowest rates of urbanization despite this large variation in average wages.\footnote{India's population is 32\% urbanized (UN, 2016). In contrast, urbanization rates are substantially higher in other economically-comparable areas such as China (60\%), Indonesia (55\%), Nigeria (50\%), and UN defined ``less developed regions'' (52\%).}

We also employ historical labor market data from the 1999-2000 round of the National Sample Survey (NSS), administered by the Government of India. This survey is also a large-scale, representative sample of the Indian population, which can then be aggregated to the district level. Using unique district and industry codes, this can be merged with the 2017-2018 PLFS data to estimate the growth and composition of wages and employment across industries in India. We use this data to construct Bartik instruments for spatial wage heterogeneity in our model of migration below.

\subsection{Supplemental survey data}

Finally, we administer a survey on Facebook among a random sample of 50,000 Indian users in 2022 (above age 18). The survey is implemented on the platform: users opt-in to the survey when prompted upon login (either through the web-based or mobile application) and give consent to share responses for research purposes. Our questionnaire captures attitudes and beliefs regarding the determinants of migration decisions and the benefits of social networks. 

We use the survey for two main purposes. First, we use hypothetical questions to capture preferences for migrating across India that are orthogonal to wages or social networks. These are used as an alternative identification approach to estimating parameters in a model of migration below. Second, we capture self-reported beliefs regarding migration and social networks to provide suggestive evidence of mechanisms behind the effect of social networks in the migration decision.

\section{Descriptive findings}
\label{sec:desc}

\begin{figure}[H]
\caption{Social Network Growth, 2014-2017}
\label{fig: network_growth}
\centering
\includegraphics[width=\textwidth]{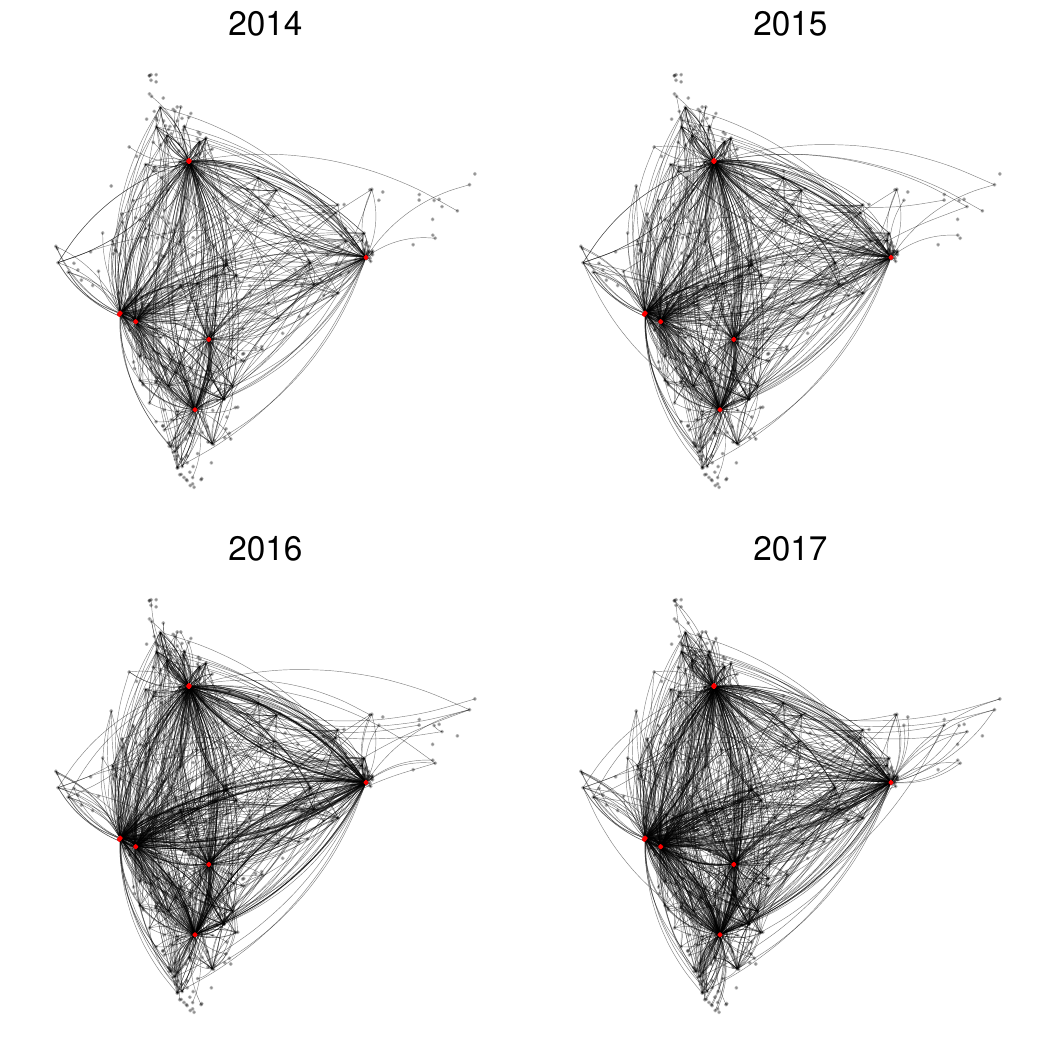}
\begin{notes}
The figure shows ties (black line segments) between any two cities (black dots) with more than 1,000 connections in our Facebook sample, at the end of each year from 2014 to 2017. Red dots depict the six most populous cities in India: Delhi, Bombay, Bangalore, Calcutta, Hyderabad, and Pune. Due to our sample construction, this reflects changes in ties between two individuals among a fixed set only, so as to not capture entry or exit of users onto the platform.
\end{notes}
\end{figure}

We present new descriptive findings on the structure of social networks and migration using our Facebook sample in India. Figure \ref{fig: network_growth} shows the spatial distribution of networks across cities in India and how they change over time, with black line segments depicting ties between any two cities with more than 1,000 connections in our Facebook sample. Between 2014 and 2017, we see that cities become increasingly more connected and particularly so in large, urban cities (depicted in red). Importantly, due to our sample construction, this reflects changes in ties between two individuals among a fixed set only, so as to not capture entry or exit of users onto the platform. Changes in this spatial network structure combine both migration patterns and social network growth. Together, these empirical patterns are consistent with three forces: (i) migration patterns are concentrated toward larger, higher-wage cities, where individuals bring their networks (wage preferences); (ii) increased networks in large cities drive further migration into them (network preferences); and (iii) social interactions at the destination expand social networks (social interactions).

\begin{figure}[H]
\caption{Migration, Social Networks, and Wages}
\label{fig: transition}
\centering
\begin{subfigure}{0.5\textwidth}
    \caption{Migration Rates}
    \label{fig: mig_wage}
    \includegraphics[width=\textwidth]{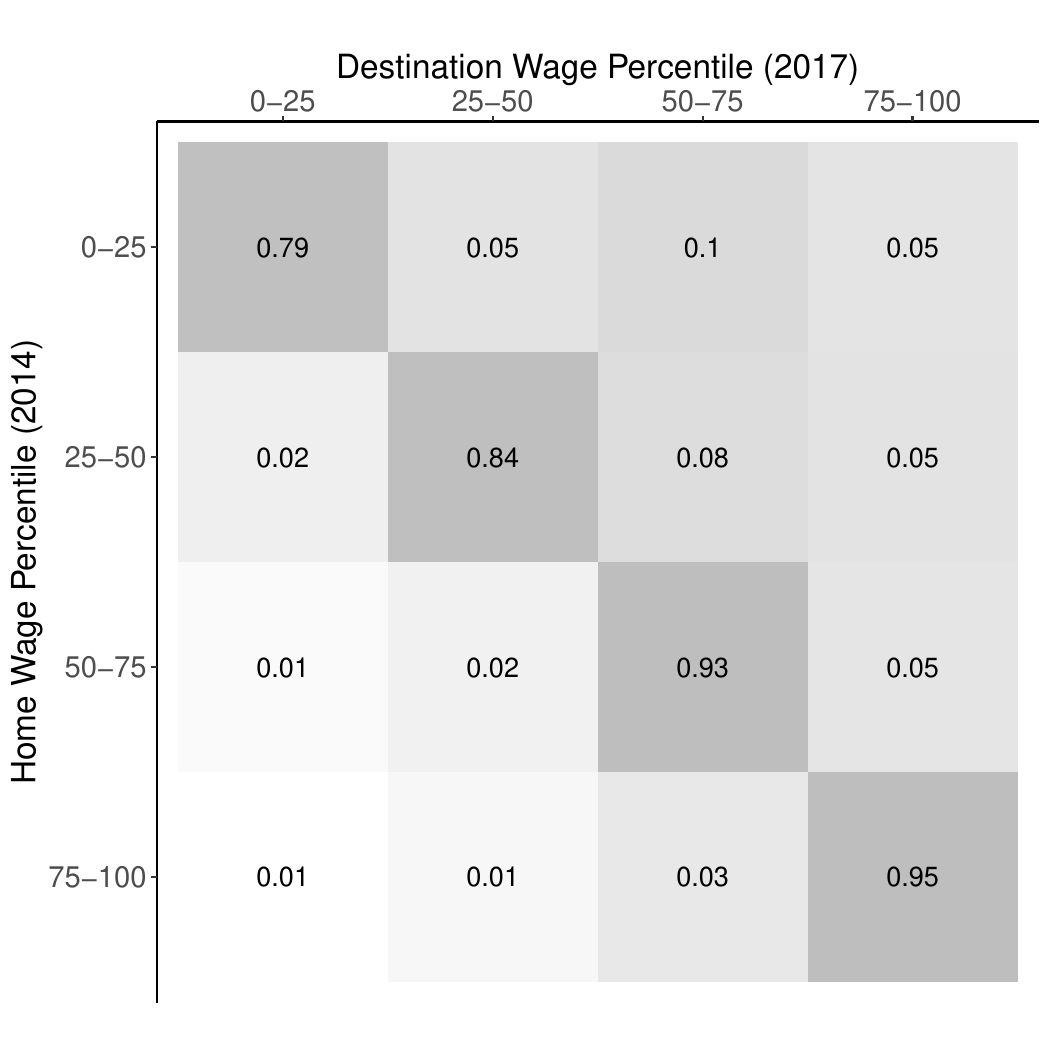}
\end{subfigure}
\hspace{-0.5em}
\begin{subfigure}{0.5\textwidth}
    \caption{Fraction of Social Ties}
    \label{fig: net_wage}
    \includegraphics[width=\textwidth]{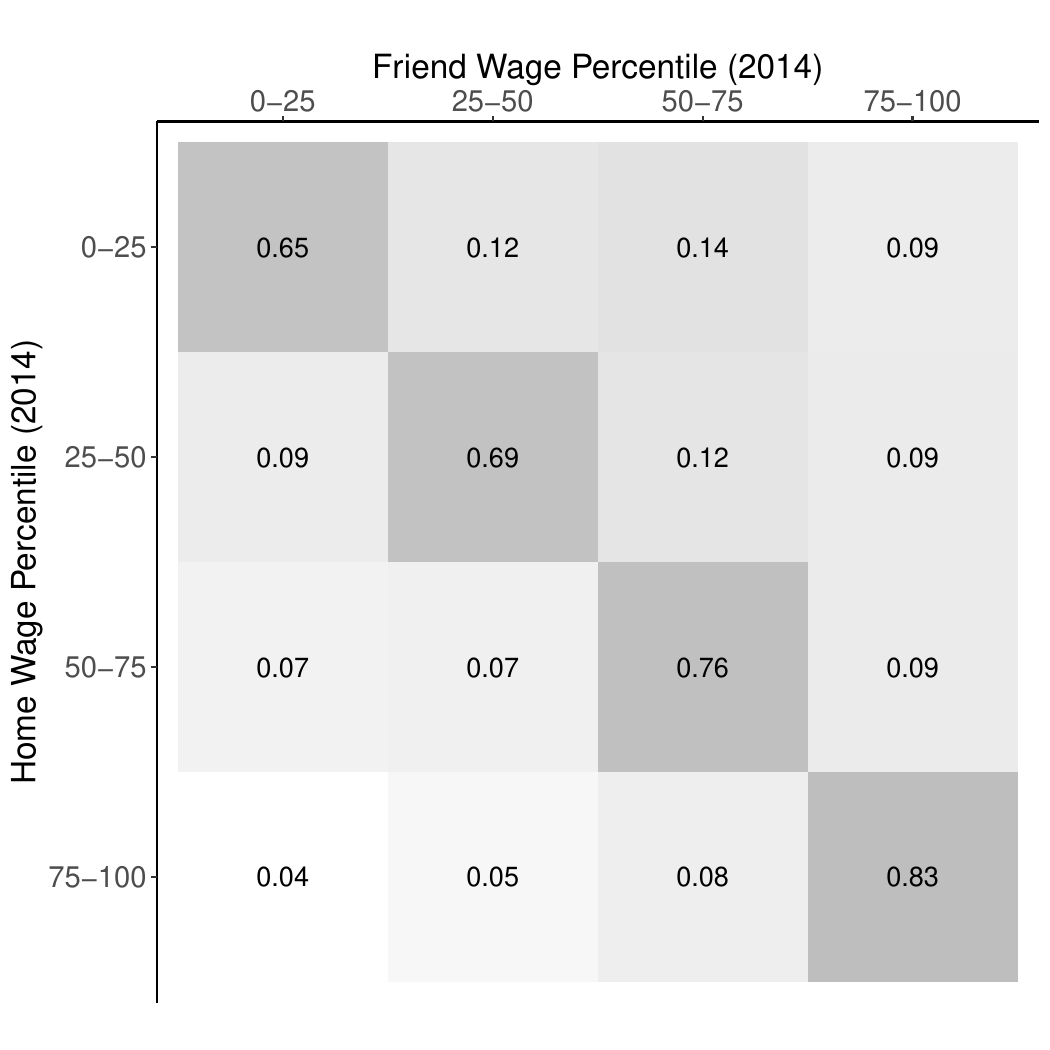}
\end{subfigure}
\begin{notes}
This figure shows transition matrices of migration and social ties by average wages across locations. Panel A reports the probability of migration from a home location in 2014 in a given percentile bin of average wages (y-axis) to a destination location in 2017 in a given percentile bin of average wages (x-axis). Panel B reports the fraction of social ties from a home location in 2014 in a given percentile bin of average wages (y-axis) to a friend location in 2014 in a given percentile bin of average wages (x-axis). Darker colors reflect higher probabilities. Wage data come from the 2017-18 round of the Periodic Labour Force Survey and all other data come from author calculations from our Facebook sample.
\end{notes}
\end{figure}

What determines these migration and social network patterns? Figure \ref{fig: transition} reports transition probabilities of migration and social ties by quartile of average wages across locations. We see that individuals have the highest probability of living in the same wage quartile from 2014 to 2017 (diagonal cells in Panel A), and have the highest concentration of social ties in the same wage quartile in 2014 (diagonal cells in Panel B). Further, compared to poorer locations, individuals are both more likely to migrate to richer locations and more likely to have friends in richer locations \emph{ex-ante}, depicted by darker cells in the northeast rather than southwest off-diagonal cells. This suggests that choices of migration across destinations correlates with both destination wages and social ties.

On average, roughly 11\% of our sample migrates out of their origin city each year. How do these migrants compare to non-migrants? Table \ref{tab: whomig} presents estimates from an OLS regression of migration status on individual characteristics. We see in column (1) that wealthier (proxied by device price), more educated, and younger individuals are more likely to migrate out of their city. However, migrants come from origin districts with lower average wages. In column (2), we include a vector of network-related observables, including the average district wage across connections, the size of the network in the origin city, and the total network size across all cities. We find that individuals with richer friends, fewer friends at origin, but more friends overall, have a higher likelihood of migration. After including these network variables, the coefficients on device price and college attendance fall by half and two-thirds, respectively, and the $R^2$ increases tenfold. This suggests first that ``social capital'' (the size and average wage of social networks) is a strong predictor of whether or not individuals migrate; and second that networks explain a substantial share of the explanatory power of wealth and human capital on migration.

\begin{table}[htbp]
\centering
\caption{Migration and Individual Characteristics}
\label{tab: whomig}
\begin{tabular}{lcc}
\toprule
Dependent Variable:&\multicolumn{2}{c}{Migrated$_t$ (pp)}\\
Model:&(1) & (2)\\
\midrule 
Log(Device Price)$_{t-1}$ & 0.6$^{***}$ & 0.3$^{**}$\\
  &(0.1) & (0.1)\\
1$\{$Attended College$\}$ & 1.5$^{***}$ & 0.5$^{*}$\\
  &(0.3) & (0.3)\\
Log(Age)$_{t-1}$ & --4.3$^{***}$ & --5.3$^{***}$\\
  &(0.3) & (0.3)\\
Log(Orig. Wage)$_{t-1}$ & --3.8$^{***}$ & --7.2$^{***}$\\
  &(0.2) & (0.3)\\
Log(Network Wage)$_{t-1}$ &    & 10.7$^{***}$\\
  &   & (0.4)\\
Log(Orig. Network Size)$_{t-1}$ &    & --8.8$^{***}$\\
  &   & (1.0)\\
Log(Network Size)$_{t-1}$ &    & 9.0$^{***}$\\
  &   & (1.0)\\
\midrule Observations & 130K&130K\\
R$^2$ & 0.006&0.066\\
\midrule\midrule\multicolumn{3}{l}{\emph{  }}\\
\end{tabular}
\begin{notes}
This table presents estimates from an OLS regression of a binary indicator of whether than individual migrates out of their residence city in 2017 on a vector of individual characteristics. This is restricted to a 1\% random sample of our Facebook sample, for which we have data on all covariates. Wage data come from the 2017-18 round of the Periodic Labour Force Survey and all other data come from author calculations from our Facebook sample. Standard errors are in parentheses and estimates are rescaled to reflect impacts on percentage points of migration rates.\\
*$p<0.10$, **$p<0.05$, ***$p<0.01$
\end{notes}
\end{table}

If an important driver of migration is social capital, which individuals have the most? Table \ref{tab: whonet} presents estimates from OLS regressions of social capital variables on individual characteristics. We find evidence that the rich, the educated, and migrants have higher social capital both in aggregate and outside of home: a 100\% increase in wealth (proxied by device price) is associated with a 35-47\% increase in network size and a 4.5-5.2\% increase in average network wages; college attendance increases network sizes by 34-37\% and network wages by 0.8-0.9\%; compared to non-migrants, migrants have 27-66\% larger network sizes and 4.1-5.2\% larger network wages.

\newpage
\begin{landscape}
\begin{table}[htbp]
\centering
\caption{Networks and Individual Characteristics}
\label{tab: whonet}
\footnotesize
\begin{tabular}{lcccc}
\toprule
Dependent Variables:&Log(Network Size)$_{t-1}$&Log(Non-Origin Network Size)$_{t-1}$&Log(Network Wage)$_{t-1}$&Log(Non-Origin Network Wage)$_{t-1}$\\
Model:&(1) & (2) & (3) & (4)\\
\midrule 
Log(Device Price)$_{t-1}$ & 0.47$^{***}$ & 0.35$^{***}$ & 0.03$^{***}$ & 0.05$^{***}$\\
  &(0.01) & (0.01) & (0.00) & (0.00)\\
1$\{$Attended College$\}$ & 0.34$^{***}$ & 0.37$^{***}$ & 0.01$^{***}$ & 0.01$^{***}$\\
  &(0.01) & (0.01) & (0.00) & (0.00)\\
Log(Age)$_{t-1}$ & --0.32$^{***}$ & 0.05$^{***}$ & 0.05$^{***}$ & 0.05$^{***}$\\
  &(0.01) & (0.01) & (0.00) & (0.00)\\
Log(Orig. Wage)$_{t-1}$ & 0.15$^{***}$ & --0.26$^{***}$ & 0.49$^{***}$ & 0.11$^{***}$\\
  &(0.01) & (0.01) & (0.00) & (0.00)\\
Migrated$_t$ & 0.27$^{***}$ & 0.66$^{***}$ & 0.04$^{***}$ & 0.05$^{***}$\\
  &(0.01) & (0.01) & (0.00) & (0.00)\\
\midrule Observations & 130K &130K&130K&130K\\
$R^2$ & 0.079&0.063&0.675&0.077\\
\midrule\midrule\multicolumn{5}{l}{\emph{  }}\\
\end{tabular}
\begin{notes}
This table presents estimates from OLS regressions of individual social capital in 2017 on a vector of individual characteristics. This is restricted to a 1\% random sample of our Facebook sample, for which we have data on all covariates. Wage data come from the 2017-18 round of the Periodic Labour Force Survey and all other data come from author calculations from our Facebook sample. Standard errors are in parentheses.\\
*$p<0.10$, **$p<0.05$, ***$p<0.01$
\end{notes}
\end{table}

\end{landscape}

\begin{figure}[H]
\centering
\caption{Survey Evidence on Determinants of Migration Choice}
\label{fig: prefcity}
\includegraphics[width=\textwidth]{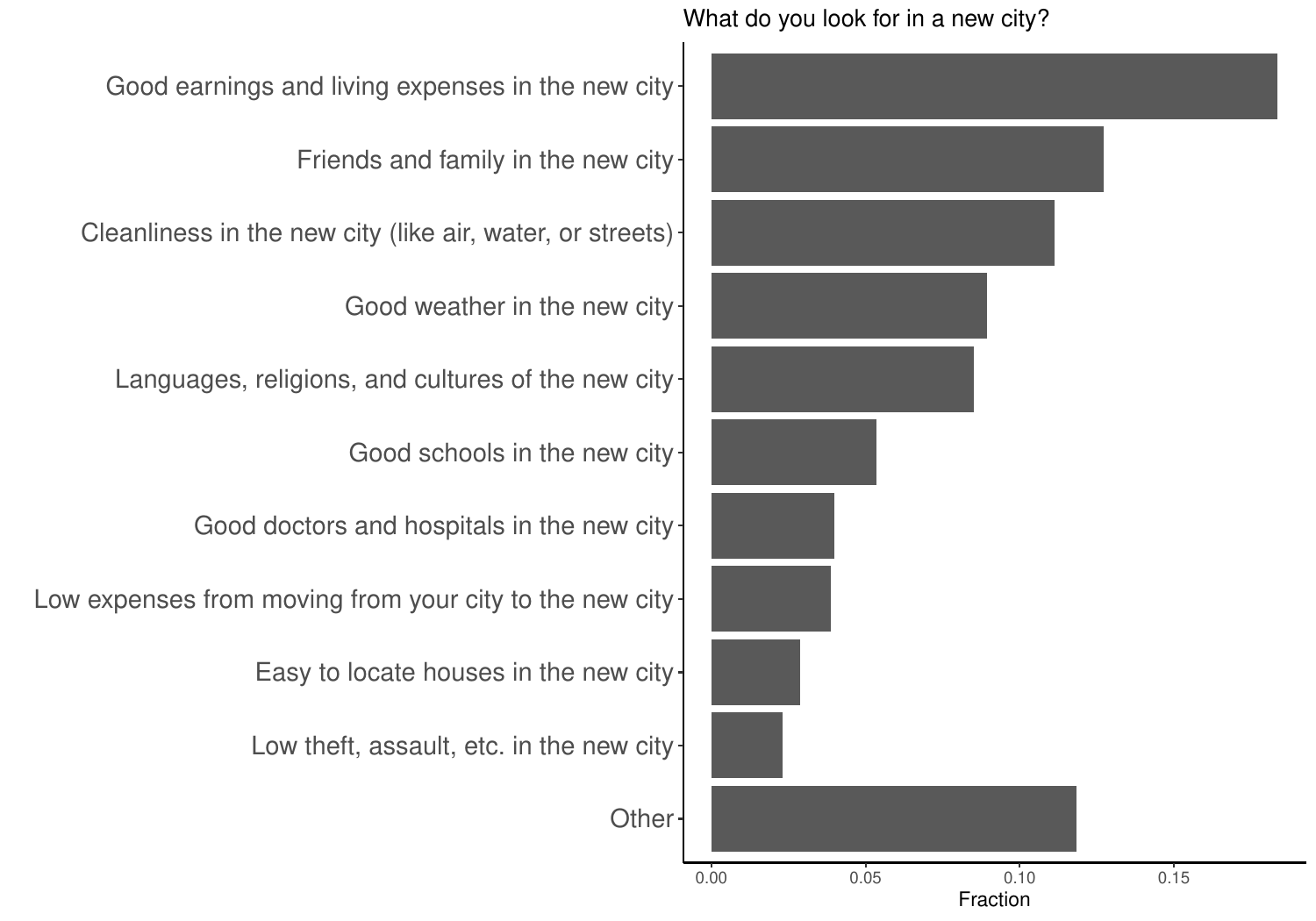}
\begin{notes}
This figure reports survey frequencies across categories for the question ``What is the most important consideration for choosing which new city to move to?'' The survey was administered among a random sample of 50,000 Indian Facebook users in 2022.
\end{notes}
\end{figure}

\newpage
To complement our observed data on migration and networks, we also administered a survey to capture beliefs regarding migration decisions. Figure \ref{fig: prefcity} reports the distribution of responses across categories to the following question:
\begin{quote}
    \emph{Imagine you are moving out of your city into a new city in India. What is the most important consideration for choosing which new city to move to?}
\end{quote}
As their most important consideration, we find that 18\% of respondents list earning and expenses and 13\% list friends and family, exceeding other amenities like sanitation, weather, education, or safety. Though self-reported, this suggests an important role for social networks in individual location decisions.

Together, these descriptive findings suggest a strong relationship between migration, social networks, and wages across locations. This motivates a model of location choice that accounts for region-specific wages, amenities, and social networks. In the next section we specify and estimate such a model that allows us the quantify the role of social networks in driving migration decisions.

\section{A model of migration}
\label{sec:model}
We model migration as a typical discrete choice problem faced by workers.\footnote{We adapt similar models of migration (labor supply) as in \cite{bm19} to incorporate region-specific social networks.} Individual $i$ maximizes utility $V_{ij}$ from living in city $j$:
\begin{equation} \label{eq:v} V_{ij} = \xi_j - \tau_{ij} + \epsilon_{ij}. \end{equation}
The term $\xi_j$ captures city-specific preferences, which can be decomposed into wage and non-wage components:
\begin{equation} 
\label{eq:wage}
\xi_j = \beta Y_j + \xi_j^A,
\end{equation}
where $Y_j$ are expected real wages in the city and $\xi_j^A$ are non-wage components of city preferences, typically unobserved (commonly referred to as ``amenities'' such as weather, environmental quality, or entertainment). 

The term $\tau_{ij}$ captures so-called ``moving costs'' that are specific to individual $i$ and destination $j$. These may include factors such as physical distance, cultural and language differences, or social networks. We parametrize these moving costs to account for both distance and social network forces:
\begin{equation}-\tau_{ij} = \gamma N_{ij} - \delta(D_{ij}),\end{equation}
where $N_{ij}$ is the (log) number of individual $i$'s social connections in city $j$, and $\delta(D_{ij})$ is some function $\delta$ of the physical distance $D_{ij}$ between individual $i$'s origin city and destination city $j$.

Finally, $\epsilon_{ij}$ captures idiosyncratic or ``match'' preferences that are specific to individual $i$ and city $j$ but are not captured in city-specific preferences $\xi_j$ or moving costs $\tau_{ij}$. For example, individuals that enjoy watching sports may particularly prefer Calcutta due to its famous cricket stadium.

The parameter $\beta$ captures the labor supply elasticity of wages -- the extent to which expected wages in the destination drive migration. The parameter $\gamma$ captures the labor supply elasticity of social networks -- the extent to which social ties in the destination drive migration. Therefore, the term $\gamma/\beta$ captures the marginal rate of substitution between changes in destination wages and destination social networks -- or in other words, the private marginal willingness-to-pay (MWTP) for social networks. More broadly, estimating these structural parameters allow us to quantify the role of social networks in driving migration, in comparison to other tangible characteristics such as wages or distance.

Traditional ``gravity'' models of migration assume moving costs increase with distance and often impose $-\tau_{ij} =  -g(D_{ij})$. In these models, $g$ captures the gravity relationship and subsumes all migrations frictions that may increase with distance including physical moving expenses, language and cultural differences, or social networks. By incorporating social networks directly into moving costs, we can compare the non-network gravity relationship $\delta$ and total gravity relationship $g$ to estimate the fraction that is explained by social networks. In addition, by allowing moving costs to depend on individual- and destination-specific networks, moving costs $\tau_{ij}$ have dramatically larger heterogeneity than what would be achieved by considering only distance (or an origin-destination pair).

It follows from the model that location choices $L_{ij}$ can be expressed as a discrete-choice problem:
\begin{equation}L_{i} = \text{arg max}_j V_{ij}.\end{equation}
Given observed location choices, wages, and networks, we can estimate this model with maximum likelihood in a traditional Logistic regression with the typical Type-1 Extreme Value distributional assumption on $\epsilon_{ij}$.\footnote{For tractability, we estimate models on a 1\% sample of our Facebook sample. Individual-year specific choice sets include all cities with at least 1 friend in period $t$ and a random sample of 10 additional cities in India. Sampling probabilities are included as a control as in \cite{m77}.}

\subsection{Identification of network effects}

There are two main threats to identification of networks effects in this model. The first is a simultaneity problem, where location decisions are determined by geographic networks, but geographic networks are themselves determined by location decisions. We address simultaneity by exploiting the panel structure of our data to estimate the effect of networks at $t-1$ on location decisions at $t$.\footnote{Similarly, \cite{bct19} include a model with lagged networks as a robustness check for this simultaneity concern.}

The second concern is networks may be endogenous. In particular, networks $N_{ij}$ may not be orthogonal to unobserved preferences $\epsilon_{ij}$. One may assume that $N_{ij}$ is orthogonal to $\epsilon_{ij}$ conditional on $\xi_j$. However, $\xi_j$ is unobserved and if $\xi_j$ and $N_{ij}$ are positively correlated, typical estimates of $\gamma$ would be biased downward. For example, networks may be large in Delhi because Delhi has high wages, or networks large in Goa because the weather in Goa is nice. 

We use three different identification approaches to address the endogeneity of networks. First, we include ``restrictive fixed effects'' as in \cite{bct19} to account for fixed preferences in narrowly-defined bins (as fine as origin-by-destination-by-year). Second, we develop an instrumental variables (IV) approach, by exploiting local productivity shocks interacted with individual network structure to deliver plausibly random variation in geographic networks. Third, we administer a survey on Facebook where we ask hypothetical questions to capture individuals' preferences for cities -- we then use these ``offline'' estimates of $\xi_j^A$ to estimate model parameters.

Below we describe (i) how we use panel data to address simultaneity when estimating the model and (ii) each identification approach we use to address network endogeneity.

\paragraph{Estimation with panel data}
To take the model to the data, we amend the general framework above to exploit the panel dimension of our data and make specific functional form assumptions. In particular, we assume decisions to migrate are made for period $t$ with information at period $t-1$. Individuals $i$ receive utility for living in city $j$ at year $t$, depending on networks and distances at year $t-1$:
\begin{equation}
\label{eq:main}
V_{ijt} =  \xi_j + \gamma N_{ij,t-1} - \delta(D_{ij,t-1}) + \epsilon_{ijt}.
\end{equation}
We interpret this assumption as limited information on others'  preferences $\epsilon_{-i,jt}$, so that individuals form expectations of networks at $t$ using current networks at $t-1$: $\mathbb{E}[N_{ijt}] = N_{ij,t-1}$. For tractability, this model is essentially static: individuals use only current draws of $\epsilon_{ijt}$ and information at $t-1$ to form migration decisions.\footnote{For tractability, we do not model dynamics. An alternative model could form expectations of $N_{ijt}$ with more sophistication, by computing expected location decisions of others $L_{-it}$ using the distribution of others' shocks $\epsilon_{-it}$ and own shock $\epsilon_{it}$. We anticipate, because of strong gravity forces in our data, this procedure would result in expectations of future networks equal to current networks. We discuss potential implications of a dynamic model in the final section.} The panel structure is used to ensure that there is no simultaneity problem: network effects on location choices at $t$ exploit variation in networks at $t-1$ only.

Finally, we model the distance-related moving cost $\delta(.)$ with fixed and variable components:
\begin{equation}\label{eq:maind} \delta(D_{ij}) = \delta_f 1\{D_{ij} \neq 0\} + \delta_v \log(D_{ij}) + \delta_{vs} \log(D_{ij})\times 1\{s(i) \neq s(j)\},\end{equation}
where $s(i)$ and $s(j)$ are the administrative states of individual $i$'s origin city and destination city $j$, respectively. With this functional form, $\delta_f$ captures the fixed cost of moving outside the city (any positive distance), $\delta_v$ captures the variable cost of moving that increases with the log of distance, and $\delta_{vs}$ captures the additional variable cost for out-of-state moves. 

This structure accounts for typical fixed and variable costs associated with moving, and also includes cultural, linguistic, and policy differences that change discontinuously across state borders in India (\cite{klmos18}). We use this panel structure and parametrization of distance across all identification strategies discussed below.

\paragraph{Approach 1: Restrictive fixed effects}
The specification of $V_{ijt}$ in equation \ref{eq:main} includes destination fixed effects $\xi_j$ to account for preferences that are fixed across individuals and time periods for each city. To address endogeneity of networks, we may model utility with even more restrictive fixed effects. For example, in the extreme case,
\begin{equation} \label{eq:fe} V_{ijt} = \gamma N_{ij,t-1} + \omega_{o(i,t-1),j,t} + \epsilon_{ijt},\end{equation}
where $o(i,t-1)$ is the origin city of individual $i$ in year $t-1$. Thus, $\omega_{o(i,t-1),j,t}$ captures fixed effects that are specific to each origin-destination-year tuple, and captures preferences that are specific to that cell. For example, $\omega_{\text{Delhi},\text{Bombay},2017}$ captures preferences that are common between all individuals that live in Delhi in 2016 and are considering moving to Bombay in 2017. Thus, estimates of $\gamma$ exploit leftover, plausibly random variation in networks after controlling for these common origin-destination-year effects.

This approach, used by \cite{bct19} to identify network effects in Rwanda, has particularly relevant properties for location preferences. Importantly, $\omega$ subsumes both preferences specific to each destination city $\xi_j$ (by summing over $o$) and moving costs that are specific to each origin-destination pair $\tau_{oj}$ such as physical distance (by summing over $t$). For comparison, we estimate the model under a variety of fixed effect specifications: no fixed effects, only destination effects $\xi_j$, destination-by-year effects $\xi_{jt}$, and origin-by-destination-by-year effects $\omega_{ojt}$.\footnote{All specifications implicitly include individual-by-year fixed effects as choices are made within each individual and year across destinations. For the model with origin-by-destination-by-year effects, we include only individual fixed effects to replicate \cite{bct19}.}

While this approach exploits the richness of data to limit the variation in networks used to identify effects, this leftover variation may still be correlated with idiosyncratic city preferences $\epsilon_{ijt}$. In the next section, we develop an instrument for networks to address endogeneity more directly.

\paragraph{Approach 2: Productivity shocks and network structure}
We seek to develop an instrument for geographic social networks that is uncorrelated with other preferences for cities. India's economy is roughly 60\% agricultural, and local productivity shocks such as droughts or heat spells may cause individuals to seek employment elsewhere. If moving costs are sufficiently high, these temporary moves may end up being permanent. Indeed, weather shocks have been shown to cause permanent, internal migration within developing countries (\cite{boh14}; \cite{mgk14}). 

While weather shocks may move individuals to other cities, they only change subsequent social networks to the extent existing individual networks overlap with these shocks. We develop an instrument for geographic social networks by exploiting quasi-random productivity shocks across space and how they interact with individuals' existing network structure.

Figure \ref{fig: instrument} shows a visual representation of such an instrument. In example 1, we show individual $i$'s network across cities, with some connected (black lines) and some unconnected. Productivity shocks are experienced by connected city $j_4$ and unconnected city $j_9$ (red circles). These shocks cause individuals to migrate to cities nearby (red arrows), $j_1$ and $j_2$ and $j_6$ and $j_7$, respectively. However, with respect to $i$'s network, migration spillovers from the connected city $j_4$ are larger (dark red arrows) compared to spillovers from the unconnected city $j_9$ (light red arrows). Thus, we expect $i$'s networks to increase more in $j_1$ and $j_2$ than $j_6$ and $j_7$. Comparing migration decisions and networks between these two pairs of destination cities allows us to isolate the effect of networks. Similarly, in example 2, we compare larger spillovers to cities $j_2$ and $j_4$ compared to city $j_8$. In example 3, because of the panel structure, we can compare migration decisions to the same city $j_8$, but between periods when shocks hit the connected nearby city $j_8$ ($t=1$) and the unconnected nearby city $j_5$ ($t=2$).

\begin{figure}[H]
    \centering
    \caption{Productivity Shocks, Migration, and Networks}
    \label{fig: instrument}
    \begin{subfigure}{0.45\textwidth}
    \caption{Example 1}
    \hspace{2em}
    \begin{tikzpicture}[scale=1]
    \Vertex[x=0,y=0,label=$i$]{A}
    \Vertex[x=-1,y=1,label=$j_4$]{B}
    \Vertex[x=-0.3,y=2,label=$j_2$]{C}
    \Vertex[x=1,y=1.2,label=$j_6$]{D}
    \Vertex[x=2.1,y=0.7,label=$j_9$]{DD}
    \Vertex[x=1.6,y=-1,label=$j_{12}$]{E}
    \Vertex[x=0.6,y=-1.2,label=$j_{10}$]{F}
    \Vertex[x=-1.3,y=-1.3,label=$j_{8}$]{G}
    \Vertex[x=-1.5,y=-0.3,label=$j_5$]{GD}
    \Vertex[x=-0.6,y=-2.1,label=$j_{14}$]{H}
    \Edge (A)(B)
    \Edge (A)(C)
    \Edge (A)(D)
    \Edge (A)(E)
    \Edge (A)(F)
    \Edge (A)(G)
    \Edge (A)(H)
    \Vertex[x=1.1,y=0.1,label=$j_7$]{I}
    \Vertex[x=1.1,y=-2,label=$j_{11}$]{J}
    \Vertex[x=-1.5,y=2,label=$j_1$]{K}
    \Vertex[x=1.25,y=2.1,label=$j_3$]{L}
    
    \Vertex[x=-1,y=1,label=$j_4$,color=red!40!white]{B}
    \Vertex[x=2.1,y=0.7,label=$j_9$,color=red!40!white]{DD}
    
    {
    \Edge[Direct,color=red!70!gray,opacity=.7,lw=1pt](B)(C)
    \Edge[Direct,color=red!70!gray,opacity=.7,lw=1pt](B)(K)
    }
   {
    \Edge[Direct,color=red!70!gray,opacity=.3,lw=1pt](DD)(D)
    \Edge[Direct,color=red!70!gray,opacity=.3,lw=1pt](DD)(I)
    }
\end{tikzpicture}
\end{subfigure}
\begin{subfigure}{0.45\textwidth}
\caption{Example 2}
    \hspace{2em}
\begin{tikzpicture}[scale=1]
    \Vertex[x=0,y=0,label=$i$]{A}
    \Vertex[x=-1,y=1,label=$j_4$]{B}
    \Vertex[x=-0.3,y=2,label=$j_2$]{C}
    \Vertex[x=1,y=1.2,label=$j_6$]{D}
    \Vertex[x=2.1,y=0.7,label=$j_9$]{DD}
    \Vertex[x=1.6,y=-1,label=$j_{12}$]{E}
    \Vertex[x=0.6,y=-1.2,label=$j_{10}$]{F}
    \Vertex[x=-1.3,y=-1.3,label=$j_{8}$]{G}
    \Vertex[x=-1.5,y=-0.3,label=$j_5$]{GD}
    \Vertex[x=-0.6,y=-2.1,label=$j_{14}$]{H}
    \Edge (A)(B)
    \Edge (A)(C)
    \Edge (A)(D)
    \Edge (A)(E)
    \Edge (A)(F)
    \Edge (A)(G)
    \Edge (A)(H)
    \Vertex[x=1.1,y=0.1,label=$j_7$]{I}
    \Vertex[x=1.1,y=-2,label=$j_{11}$]{J}
    \Vertex[x=-1.5,y=2,label=$j_1$]{K}
    \Vertex[x=1.25,y=2.1,label=$j_3$]{L}
    
     {
        \Vertex[x=-0.6,y=-2.1,label=$j_{14}$,color=red!40!white]{H}
        \Vertex[x=-1.5,y=2,label=$j_1$,color=red!40!white]{K}
        \Edge[Direct,color=red!70!gray,opacity=.7,lw=1pt](H)(G)
        \Edge[Direct,color=red!70!gray,opacity=.3,lw=1pt](K)(B)
        \Edge[Direct,color=red!70!gray,opacity=.3,lw=1pt](K)(C)
    }
\end{tikzpicture}
\end{subfigure}
\vspace{0.5em}
\begin{subfigure}{\textwidth}
\caption{Example 3}
\begin{subfigure}{0.45\textwidth}
    \hspace{3.5em}
  \begin{tikzpicture}[scale=1]
    \Vertex[x=0,y=0,label=$i$]{A}
    \Vertex[x=-1,y=1,label=$j_4$]{B}
    \Vertex[x=-0.3,y=2,label=$j_2$]{C}
    \Vertex[x=1,y=1.2,label=$j_6$]{D}
    \Vertex[x=2.1,y=0.7,label=$j_9$]{DD}
    \Vertex[x=1.6,y=-1,label=$j_{12}$]{E}
    \Vertex[x=0.6,y=-1.2,label=$j_{10}$]{F}
    \Vertex[x=-1.3,y=-1.3,label=$j_{8}$]{G}
    \Vertex[x=-1.5,y=-0.3,label=$j_5$]{GD}
    \Vertex[x=-0.6,y=-2.1,label=$j_{14}$]{H}
    \Edge (A)(B)
    \Edge (A)(C)
    \Edge (A)(D)
    \Edge (A)(E)
    \Edge (A)(F)
    \Edge (A)(G)
    \Edge (A)(H)
    \Vertex[x=1.1,y=0.1,label=$j_7$]{I}
    \Vertex[x=1.1,y=-2,label=$j_{11}$]{J}
    \Vertex[x=-1.5,y=2,label=$j_1$]{K}
    \Vertex[x=1.25,y=2.1,label=$j_3$]{L}
     \Vertex[x=-0.6,y=-2.1,label=$j_{14}$,color=red!40!white]{H}
    \Edge[Direct,color=red!70!gray,opacity=.7,lw=1pt](H)(G)
    \node at (0,-3) {$t=1$};
\end{tikzpicture}
\end{subfigure}
\begin{subfigure}{0.45\textwidth}
    \hspace{3.5em}
\begin{tikzpicture}[scale=1]
    \Vertex[x=0,y=0,label=$i$]{A}
    \Vertex[x=-1,y=1,label=$j_4$]{B}
    \Vertex[x=-0.3,y=2,label=$j_2$]{C}
    \Vertex[x=1,y=1.2,label=$j_6$]{D}
    \Vertex[x=2.1,y=0.7,label=$j_9$]{DD}
    \Vertex[x=1.6,y=-1,label=$j_{12}$]{E}
    \Vertex[x=0.6,y=-1.2,label=$j_{10}$]{F}
    \Vertex[x=-1.3,y=-1.3,label=$j_{8}$]{G}
    \Vertex[x=-1.5,y=-0.3,label=$j_5$]{GD}
    \Vertex[x=-0.6,y=-2.1,label=$j_{14}$]{H}
    \Edge (A)(B)
    \Edge (A)(C)
    \Edge (A)(D)
    \Edge (A)(E)
    \Edge (A)(F)
    \Edge (A)(G)
    \Edge (A)(H)
    \Vertex[x=1.1,y=0.1,label=$j_7$]{I}
    \Vertex[x=1.1,y=-2,label=$j_{11}$]{J}
    \Vertex[x=-1.5,y=2,label=$j_1$]{K}
    \Vertex[x=1.25,y=2.1,label=$j_3$]{L}
    \Vertex[x=-1.5,y=-0.3,label=$j_5$,color=red!40!white]{GD}
    \Edge[Direct,color=red!70!gray,opacity=.3,lw=1pt](GD)(G)
    \node at (0,-3) {$t=2$};
\end{tikzpicture}
\end{subfigure}
\end{subfigure}
\vspace{0.5em}
\begin{notes}
This figure depicts visual intuition for the instrument. Each panel is one example of how weather shocks interact with individual networks. Each bubble is a city $j_1,\dots,j_{12}$. The central bubble denoted by $i$ is individual $i$'s origin city. Lines indicate whether or not individual $i$ is connected to the respective city. Red bubbles denote cities that experience weather shocks, and arrows indicate the migration spillovers from those cities. Darker (lighter) arrows indicate larger (smaller) migration spillovers of $i$'s friends, given that shocked cities are connected (not connected) to $i$. The variation we exploit is comparing cities that receive these larger versus smaller spillovers from connected versus unconnected shocked cities. Panels A and B provide examples across cities for a given time period, and Panel B shows an example across time periods for a given city.
\end{notes}
\end{figure}

In order to construct an instrument for networks at $t-1$, we use weather shocks at $t-2$ which, given the model in equation \ref{eq:main}, affect location decisions one period later. Because location choices are a function of distance, we expect those that experience weather shocks to migrate nearby rather than far away. We define $w(j,t)$ as the city closest to city $j$ that experiences a weather shock at time $t$:
\begin{equation} w(j,t) = \text{argmin}_{j' \in J_t^{\text{Shock}}} D_{jj'}, \end{equation}
where $J_t^{\text{Shock}}$ is the set of cities in period $t$ are are hit by a weather shock. Thus, networks $N_{ij,t-1}$ depend on prior weather shocks nearby $w(j,t-2)$. In particular, shocks impact networks in $j$ depending on $j$'s distance to the shocked city $D_{j,w(j,t-2)}$ and the number of connections in the shocked city $N_{i,w(j,t-2),t-2}$. This suggests the following first-stage regression:

\begin{equation} 
\label{eq:fs}
N_{ij,t-1} = \theta_1 N_{i,w(j,t-2),t-2} + \delta(D_{j,w(j,t-2)}) + g(X_{ij,t-1}) + \tilde{\epsilon}_{ij,t-1},
\end{equation}
where $\delta(D_{j,w(j,t-2)})$ controls for distance to the nearest shocked city as in equation \ref{eq:maind}, $h(X_{ij,t-1})$ includes controls from the main model in equation \ref{eq:main}, and $\tilde{\epsilon}_{ij,t-1}$ is a residual.\footnote{Controls in $h$ include $\delta(D_{ij,t-1})$ and destination and individual-by-year fixed effects.} Our instrument is therefore $N_{i,w(j,t-2),t-2}$, which captures the extent to which individual $i$ is connected to a city near to $j$ that experienced a weather shock in period $t-2$. 

We argue that the instrument $N_{i,w(j,t-2),t-2}$ -- the number of connections to the nearest shocked city -- satisfies classic IV assumptions. For exogeneity, conditional on destination and individual-by-year fixed effects, city-level shocks to weather are quasi-random and orthogonal to unobserved determinants of individual social networks or migration patterns. Thus, because the set of shocked cities $J_t^{\text{Shock}}$ is conditionally quasi-random, the nearest shocked city $w(j,t)$ as well as the number of connections to $w(j,t)$ is also quasi-random. For exclusion, nearby weather shocks alone may impact migration in ways other than through social networks, including local economic spillovers and locally correlated weather. However, we argue that the extent to which individuals are connected to nearby shocked cities -- and not the shock to the city itself -- only impacts migration patterns through subsequent social networks.

Given the literature on climate-induced migration, we construct instruments based on two types of productivity shocks: low rainfall (drought) and high temperature (heat shock). In particular, we define a city-year as experiencing a ``drought'' if annual rainfall for that city is below the 15th percentile in the long-run distribution of annual rainfall for that city (1980 to 2017). Similarly, we denote a ``heat shock'' if the number of days per year exceeding 95$^\circ$F falls above the 85th percentile in the long-run distribution. Historical, grid-level weather data come from the commonly used University of Delaware (U-DEL) database. Grids are matched to cities using cities' geographical centroids.

For intuition, we report estimates from a simplified version of the first-stage regression in equation \ref{eq:fs}:
\begin{align}
    N_{ij,t-1} &= \theta_1 1\{N_{i,w(j,t-2),t-2}>0\} + \delta_f^w 1\{D_{j,w(j,t-2)}=0\} + \delta_v^w \log(D_{j,w(j,t-2)}) \notag \\
    & + \delta_{vn}^w 1\{N_{i,w(j,t-2),t-2}>0\} \times \log(D_{j,w(j,t-2)}) + g(X_{ij,t-1}) + \tilde{\epsilon}_{ij,t-1}. \label{eq:fs_simple}
\end{align}
Here, $\theta_1$ captures the first-stage effect: the additional spillover to $i$'s network in city $j$ in period $t-1$ if $i$ has at least 1 friend in the closest shocked city to $j$ in period $t-2$. The term $\delta_f^w$ captures the effect of a weather shock in city $j$ in period $t-2$ (if the distance to the shocked city is zero) on friends in $j$ in period $t-1$. The term $\delta_v^m$ captures the spillover effect that changes with distance to the shocked city. Finally, the term $\delta_{vn}^w$ captures how the spillover additionally changes with distance if $i$ has at least 1 friend in the shocked city in period $t-2$.

\begin{figure}[H]
    \centering
    \caption{First Stage: Spillovers of Nearby Weather Shocks}
    \label{fig: fs_curve}
    \includegraphics[width=\textwidth]{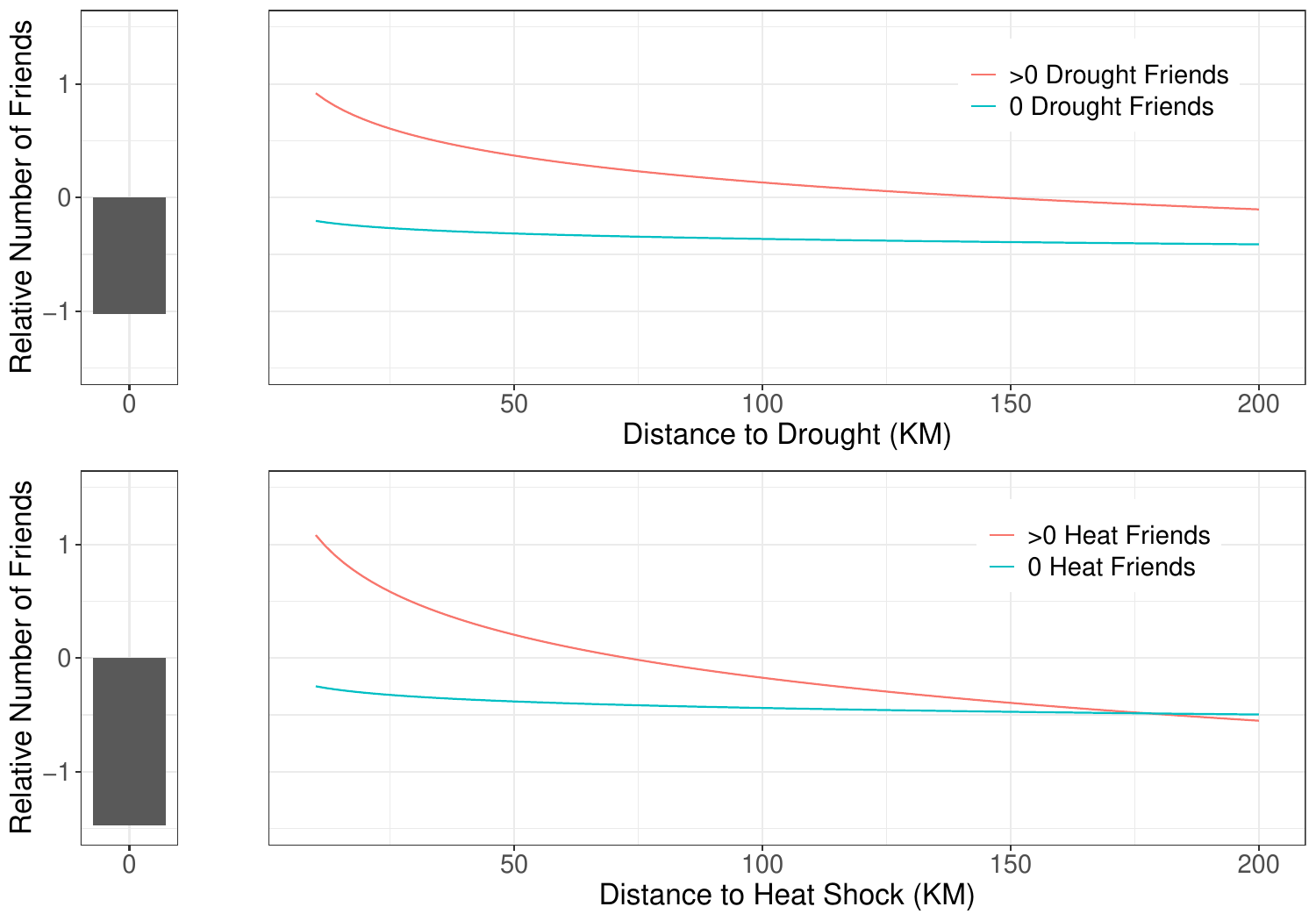}
    \begin{notes}
    This figure plots results from the simplified first stage specification in equation \ref{eq:fs_simple}, across two weather-based instruments: droughts (top) and heat shocks (bottom). Lines plot predicted number of friends in city $j$ in period $t-1$ as a function of the distance from city $j$ to the nearest city that experienced a weather shock in period $t-2$. Blue lines show estimates the effect for shocked cities with no friends in period $t-2$ and red lines show the effect for socked cities with at least 1 friend. Bars on the left depict the effect of zero distance: the effect of city $j$ experiencing a weather shock in period $t-2$ on number of friends in city $j$ in period $t-1$. Effects on networks in $j$ are relative to having zero friends in the shocked city, which is just outside city $j$.
    \end{notes}
\end{figure}

Figure \ref{fig: fs_curve} plots predicted friends in city $j$ in period $t-1$ as a function of distance to the weather shock in period $t-2$ using estimates from the simplified first-stage, for both droughts (top) and heat shocks (bottom). The bars on the left show that droughts and heat shocks in city $j$ deliver roughly a 1 and 1.5 friend decline in $i$'s networks in $j$ in the subsequent year ($\delta_f^w<0$).\footnote{We find that these weather-shock migration effects fall with urbanity at the destination, consistent with a local productivity shock mechanism.} As shown in the red and blue lines, spillovers from these shocks then decline with distance from the shocks, showcasing how individuals exhibit gravity forces and migrate from shocked cities to closer destinations ($\delta_v^w<0$). The declines in distance are steeper if $i$ has at least 1 friend in the closest shocked city ($\delta_{vn}^w<0$). Importantly, at all distances, spillovers are larger when $i$ has at least 1 friend in the closest shocked city. This vertical difference between the red and blue curves depicts this variation in networks delivered by our instrument ($\theta_1>0$).

We estimate the effect of networks on migration $\gamma$ by employing a control function approach to IV estimation as in \cite{t09}, where residuals from the first-stage in equation \ref{eq:fs} are included as controls in the main model in equation \ref{eq:main}. We interpret these estimates of $\gamma$ as Local Average Treatment Effects (LATEs) for specific groups of compliers: those whose networks change as a result of local spillovers from weather shocks, that would not change otherwise. In particular, networks that are changing are driven by weather-based migration, presumably by individuals that are fleeing from local negative productivity shocks. These individuals are thus not a random sample of social networks, and so in the presence of network effect heterogeneity, effects for the average friend may differ. For example, the networks that are impacted by the instrument may be relatively more rural (employed by agriculture) and have lower incomes. In this case, if migration is more responsive to higher income social connections, we can interpret these LATE estimates of $\gamma$ as a lower bound.

\paragraph{Approach 3: Eliciting preferences from hypothetical survey questions}

The main threat to identification of networks effects is endogeneity: in particular, that networks $N_{ij}$ are correlated with unobserved, non-wage city preferences $\xi_j^A$ (``amenities''). We may assume that conditional on $\xi_j^A$, networks $N_{ij}$ are exogenous. However, because $\xi_j^A$ is unobserved, joint estimation with $\gamma$ may produce biased estimates. 

In our third identification approach, we tackle this by estimating $\xi_j^A$ directly, holding $N_{ij}$ fixed. We do so by administering hypothetical survey among 50,000 Indian Facebook users and asking where they would want to live if networks or wages were not a concern.\footnote{This is similar to other identification approaches in the empirical industrial organization literature. For example, \cite{dno20} administer hypothetical questions on preferences for private schools holding prices fixed, to estimate school-specific unobservables and identify price elasticities.} In particular, we ask individuals:
\begin{quote}
    \emph{Suppose you had the same friends, family, earnings, and expenses no matter where you lived. In this case, where in India would you want to live?}
\end{quote}
Given the hypothetical question, we assume individuals choose locations to maximize the adjusted utility without network and wage effects ($\gamma=0,\beta=0$):
\begin{equation} \tilde{V}_{ij} = \xi_j^A + \delta(D_{ij}) + \epsilon_{ij}. \end{equation}
Thus, using choices of ``dream'' cities from the survey, we can estimate $\xi_j^A$ using a typical Logit estimation. We then include these first-step estimates $\hat{\xi}_j^A$ into the main model of migration choice to estimate $\gamma$. Conditional on $\xi_j^A$, we assume variation in $N_{ij}$ is now exogenous, removing the need for instruments.

\begin{figure}[H]
    \centering
    \caption{Net Differences: Dream vs. Current}
    \label{fig: net_surv}
    \includegraphics[width=0.75\textwidth]{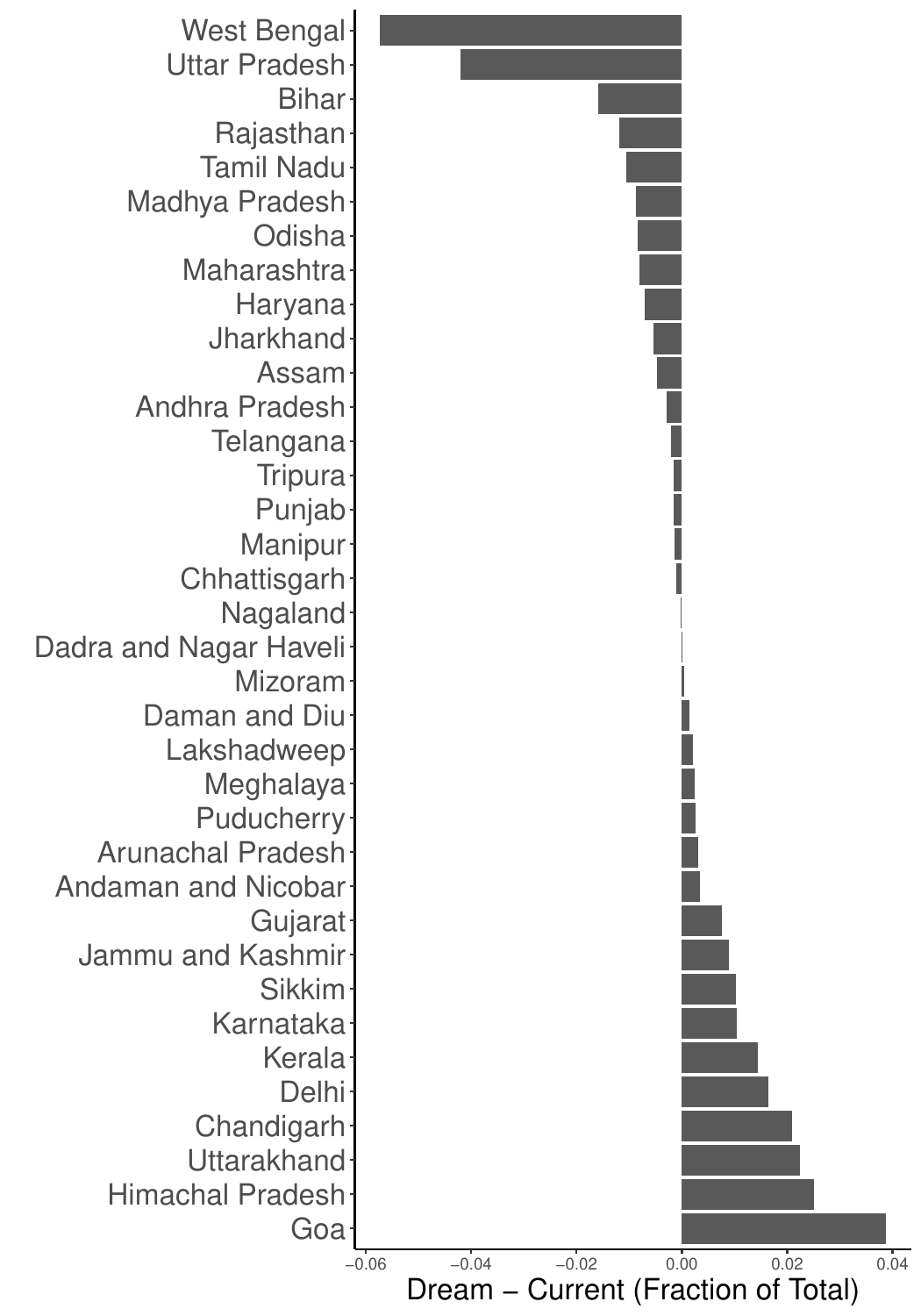}
    \begin{notes}
    This figure presents data from hypothetical survey questions. Users were asked their current city of residence as well as their ``dream'' city: where they would live if social networks or wages were not a concern. For any given state (y-axis), the bar denotes the difference in frequency between listing the dream city in the state and listing the current city in the state (x-axis). The survey was administered among a random sample of 50,000 Indian Facebook users in 2022.
    \end{notes}
\end{figure}

Figure \ref{fig: net_surv} plots hypothetical net migration across states if individuals moved from their current city to their ``dream'' city if networks and wages were not a concern. Roughly 60\% of respondents would chose to live outside their residence city if these constraints were lifted. Locations with large hypothetical net migrations would have large amenity values $\xi_j^A$. We find that individuals would move from states like West Bengal and Uttar Pradesh to states like Goa and Himachal Pradesh. The modal dream city is in Goa, a popular vacation destination known for good weather and beaches. Goa would thus have a high estimated $\xi_j^A$ from the model, consistent with intuition. While we caution that this survey data is hypothetical and self-reported (compared to revealed-preference data used in the other approaches), we conclude that amenity estimates from the survey are reasonable and in line with intuition.

\subsection{Identification of wage effects}
While the literature on network effects in migration is limited, we borrow from the spatial economics literature to estimate the effects of wages in the migration decision in equation \ref{eq:wage}. The typical endogeneity concern is that wages $Y_j$ are correlated with unobserved amenities $\xi_j^A$, such that estimates of the wage effect $\beta$ are biased. In spatial models, we may expect low-wage locations to have high amenities to compensate workers for living there and ensure equilibrium (\cite{roback82}). In this case, $\beta$ would be downward biased absent an instrument.

We use historical data on local labor markets in India to develop traditional Bartik instruments for spatial wage heterogeneity. Consistent with past work, we exploit heterogeneity in historical local industry composition (shares) together with industry-specific long-run national trends (shocks).\footnote{Past work uses similar Bartik instruments to identify wage effects in the labor supply decision in spatial equilibrium models. See, for example, \cite{bgs18}, \cite{klms21}, \cite{mrr18}, \cite{mo18}, and \cite{p20}.} We construct two such instruments using changes in wages from 1999 to 2016 weighted by labor shares or changes in labor weighted by wage shares:
\begin{equation}\label{eq:b1}B_j^{\Delta Wage} = \sum_{k} \frac{L_{jk}^{1999}}{L_j^{1999}} \Delta \overline{Y}_{k}^{1999-2016}, \end{equation}
\begin{equation}\label{eq:b2}B_j^{\Delta Labor} = \sum_{k} \frac{Y_{jk}^{1999}}{Y_j^{1999}} \Delta \overline{L}_{k}^{1999-2016},\end{equation}
where, for workers in city $j$ in industry $k$ in year $t$, $L_{jk}^t$ is the number of workers and $Y_{jk}^t$ is the average wage; $L_j^t = \sum_k L_{jk}^t$ and $Y_j^t = \sum_k Y_{jk}^t$; $\Delta \overline{Y}_{k}^{t_1-t_0} =  \overline{Y}_{k}^{t_1} - \overline{Y}_{k}^{t_0}$ and $\Delta \overline{L}_{k}^{t_1-t_0} =  \overline{L}_{k}^{t_1} - \overline{L}_{k}^{t_0}$; and $\overline{Y}_{k}^{t} = \sum_j{Y_{jk}^t}/|J|$ and $\overline{L}_{k}^{t} = \sum_j{L_{jk}^t}/|J|$.\footnote{While location choices in the main migration model are at the city level, wages are only observed at the district-level. Nonetheless, wage heterogeneity is large with over 300 district boundaries across 13 industry codes consistently linked from 1999 to 2016.}

To estimate $\beta$, we perform a two-step procedure: first, we estimate $\xi_j$ from the main migration model in equation \ref{eq:main}, which includes both wage and non-wage components; second, we estimate $\beta$ from a 2SLS specification using the first-step estimates $\hat{\xi}_j$, wages $Y_j$, and instruments $B_j \in \{B_j^{\Delta Wage},B_j^{\Delta Labor}\}$:
\begin{equation}\hat{\xi}_j = \beta Y_j + \xi^A_j, \end{equation}
\begin{equation} Y_j = \beta^{FS}B_j + \epsilon_j^{FS}.\end{equation}
Here, we assume that our instruments $B_j$ are exogenous and excluded, impacting fixed city preferences only through wages. This assumption comes from the fact that instruments exploit the historical (nearly two decades prior) labor market composition across industries and industry-specific growth at the national level, presumably orthogonal to current, city-specific preferences. We estimate effects across both instruments and include city-specific controls for robustness.

\section{The effects of networks on migration}
\label{sec:modelresults}
Table \ref{tab: glm_compare_surv} reports model estimates across identification strategies. Columns (1-4) report estimates from restrictive fixed effects, with increasingly granular fixed effect bins. Columns (5-6) report estimates using instruments constructed from drought and heat shocks, respectively. Column 7 reports estimates using the hypothetical survey approach, with amenities estimated in the first step and included in the main migration model along with wages in the second step. 

First, we find that, across specifications, migration exhibits strong gravity forces ($\delta_f>0,\delta_v<0,\delta_{vs}>0$). Across destinations, individuals prefer moving to cities that are closer (or not moving at all), and the distance effect is roughly 10\% smaller for in- vs. out-of-state migrations (for almost every specification). This suggests moving costs change discontinuously at state borders, consistent with policy, linguistic, or cultural differences (\cite{klmos18}).

Second, we find that destination social networks are important determinants of migration choice. We estimate a large marginal willingness-to-pay (MWTP) for networks in terms of distance: individuals are indifferent between a 1\% increase in destination social networks and a 0.70-1.33\% decrease in total distance. Importantly, while changing the granularity of fixed effects has only a modest impact on network effects, the IV approach reduces the MWTP for networks by 24-47\% compared to restrictive fixed effects. The survey approach also reduces MWTP by 16-26\%. This suggests that network effects may be upward biased, consistent with the idea that the geographic composition of social networks are positively correlated with city-specific preferences. Correcting for this endogeneity (with either network instruments or separately estimated amenities) reduces the effect of networks. For the remaining analysis, we use the specification with the smallest network effect (the heat shock IV approach in column 6, MWTP=0.7), and interpret resulting estimates as a lower bound.

\newpage
\begin{landscape}
\begin{table}[htbp]
\footnotesize
\centering
\caption{Model Estimates Across Specifications}
\label{tab: glm_compare_surv} 
\begin{tabular}{lccccccc}
\toprule
Dependent Variable:&\multicolumn{7}{c}{Choice$_t$}\\
  & No $\xi$ & Baseline & Time Varying $\xi$ & BCT & Drought IV & Heat IV & Survey Moments\\
Model:&(1) & (2) & (3) & (4) & (5) & (6) & (7)\\
\midrule 
Same City$_{t-1}$ & 2.10$^{***}$ & 3.69$^{***}$ & 4.33$^{***}$ &    & 3.88$^{***}$ & 4.05$^{***}$ & 1.93$^{***}$\\
  &(0.05) & (0.09) & (0.10) &    & (0.10) & (0.10) & (0.06)\\
Log(Distance)$_{t-1}$ & --0.43$^{***}$ & --0.25$^{***}$ & --0.17$^{***}$ &    & --0.32$^{***}$ & --0.36$^{***}$ & --0.57$^{***}$\\
  &(0.01) & (0.02) & (0.02) &    & (0.02) & (0.02) & (0.01)\\
Log(Distance)$_{t-1}$ $\times $ Same State$_{t-1}$ & 0.01 & 0.02$^{**}$ & 0.03$^{***}$ &    & 0.03$^{***}$ & 0.04$^{***}$ & --0.02$^{**}$\\
  &(0.01) & (0.01) & (0.01) &    & (0.01) & (0.01) & (0.01)\\
Log(Dest. Friends)$_{t-1}$ & 0.86$^{***}$ & 1.00$^{***}$ & 1.02$^{***}$ & 1.02$^{***}$ & 0.79$^{***}$ & 0.65$^{***}$ & 0.84$^{***}$\\
  &(0.01) & (0.01) & (0.01) & (0.03) & (0.04) & (0.04) & (0.01)\\
Log(Dest. Wage) &    &    &    &    &    &    & 0.72$^{***}$\\
  &    &    &    &    &    &    & (0.03)\\
\midrule MWTP for Networks (Distance) & 1.18 & 1.30 & 1.33 & 1.24 & 0.91 & 0.70 & 0.99\\
\midrule \emph{Fixed-effects}&   &   &   &   &   &   &  \\
Indiv. $\times$ Year & Y & Y & Y &  & Y & Y & Y\\
Dest. &  & Y &  &  & Y & Y & \\
Dest. $\times$ Year &  &  & Y &  &  &  & \\
Orig. $\times$ Dest. $\times$ Year &  &  &  & Y &  &  & \\
Individual &  &  &  & Y &  &  & \\
\midrule 
Observations & 750K &750K&750K&750K&750K&750K&750K\\
\midrule\midrule\multicolumn{8}{l}{\emph{  }}\\
\end{tabular}
\begin{notes}
This table reports parameter estimates across specifications and identification strategies of a discrete-choice model of migration across destination cities in period $t$ on distance and networks in period $t-1$ as in equation \ref{eq:main}. Columns (1-4) report estimates from the ``restrictive fixed effects'' approach in equation \ref{eq:fe}, with increasingly granular fixed effects. Column 4 replicates the specification in \cite{bct19} (``BCT'') with origin-by-destination-by-year effects. Columns (5-6) report estimates from the instrument approach, using drought and extreme heat shocks, respectively. We employ a control function approach to IV estimation as in \cite{t09}, where residuals from the first-stage in equation \ref{eq:fs} are included as controls in the main model in equation \ref{eq:main}. Column 7 reports estimates from the hypothetical survey question approach, where we (1) estimate amenities $\xi_j^A$ using survey choices and (2) include these estimated $\hat{\xi}_j^A$ and wages in the main migration model in equation \ref{eq:main}. Estimates of ``MWTP for Networks'' across specifications come from dividing the coefficient on Log(Dest. Friends) ($\gamma$) by the implied coefficient on Log(Distance) ($\delta$). We estimate $\delta$ by regressing moving costs from each model (predicted utility using only distance terms or origin-by-destination-by-year effects) on Log(Distance) only. Estimates are restricted to a 1\% sample of users from 2014 to 2017 over which all specifications can be estimated, and individual-year specific choice sets include all cities with at least 1 friend in period $t$ and a random sample of 10 additional cities. Sampling probabilities are included as a control as in \cite{m77}. Standard errors are in parenthesis and clustered at the individual-by-year level.\\
*$p<0.10$, **$p<0.05$, ***$p<0.01$
\end{notes}
\end{table}

\end{landscape}

\subsection{The effects of wages}
Table \ref{tab: bartik} reports results from our two-step approach to estimating the wage elasticity. First, we estimate unobserved, city-specific preferences $\xi_j$ from our preferred specification using the heat shock instrument for network effects. Then, we estimate the effect of destination wages on these estimated city-specific preferences $\hat{\xi}_j$. For comparison, we estimate $\beta$ across OLS and IV approaches using our bartik instruments in equations \ref{eq:b1}-\ref{eq:b2}.

We find first that individuals strongly prefer moving to cities with high average wages. This allows us to quantify network effects in terms of wages, as individuals are trading off potential wage gains for larger networks in equilibrium. We find that individuals are indifferent between a 1\% increase in destination networks and a 0.5-0.8\% increase in destination wages, suggesting that while the wage effect is larger (in percentage terms) than either distance or networks, network effects are also strong in monetary terms.

We also find that this wage elasticity is downward biased: moving from column (1) to (5), after including controls or constructed instruments, the effect of wages rises and the MWTP for networks falls. This suggests a negative relationship between wages and non-wage city-specific preferences (``amenities''). This is consistent with a model of compensating differentials in which cities have larger amenities to compensate workers for low wages or larger wages to compensate for lower amenities (\cite{roback82}). In addition, we find that the choice of instrument only marginally changes estimates. For the remaining analysis, we use the estimate that includes both instruments as our preferred specification for the wage elasticity (column 5).

\begin{table}[H]
\centering
\caption{The Wage Elasticity of Migration}
\label{tab: bartik}
\begin{tabular}{lccccc}
\toprule
Dependent Variable:&\multicolumn{5}{c}{Dest. $\hat{\xi}$}\\
  & \multicolumn{2}{c}{OLS} & \multicolumn{3}{c}{IV}\\
Model:&(1) & (2) & (3) & (4) & (5)\\
\midrule Log(Dest. Wage) & 0.79$^{***}$ & 0.82$^{***}$ & 1.20$^{***}$ & 1.13$^{***}$ & 1.17$^{***}$\\
  &(0.09) & (0.09) & (0.15) & (0.11) & (0.14)\\
\midrule Additional Controls & N & Y & Y & Y & Y\\
MWTP for Networks (Wages) & --0.81 & --0.78 & --0.53 & --0.57 & --0.55\\
Bartik Instruments &  &  & $\Delta Wage$ & $\Delta Labor$ & Both\\
\midrule \emph{Fixed-effects}&   &   &   &   &  \\
Indiv. $\times$ Year & Y & Y & Y & Y & Y\\
\midrule Observations & 7.5M &7.5M&7.5M&7.5M&7.5M\\
Adjusted R$^2$ & 0.185&0.228&0.201&0.209&0.204\\
F-test (1st stage) & &&7.5M&2.7M&4.4M\\
\midrule\midrule\multicolumn{6}{l}{\emph{  }}\\
\end{tabular}
\vspace{-1em}
\begin{notes}
This table reports parameter estimates across specifications of the effect of destination wages and estimated amenities from the preferred specification (Column 6 of Table \ref{tab: glm_compare_surv}). Columns (1-2) report estimates from OLS, with and without additional destination controls. Columns (3-4) report IV estimates using the constructed bartik instruments $B_j^{\Delta Wage}$ and $B_j^{\Delta Labor}$ in equations \ref{eq:b1}-\ref{eq:b2}. Finally, column (5) reports IV estimates using both bartik instruments. Estimates of ``MWTP for Networks'' across specifications come from dividing the coefficient on Log(Dest. Friends) ($\gamma$) by the coefficient on Log(Dest. Wage) ($\beta$). Estimates are restricted to a 1\% sample of users from 2014 to 2017 over which all specifications can be estimated, and individual-year specific choice sets include all cities with at least 1 friend in period $t$ and a random sample of 10 additional cities. Sampling probabilities are included as a control as in \cite{m77}. Individual-by-year fixed effects are included for all specifications and standard errors are in parentheses and clustered at the individual-by-year level.\\
*$p<0.10$, **$p<0.05$, ***$p<0.01$
\end{notes}
\end{table}

\subsection{Gravity heterogeneity by networks}
We find that, in general, the marginal cost of distance is smaller for in-state destinations, suggesting heterogeneity in the gravity relationship. We extend our simple model of migration to account for heterogeneity in preferences for distance by networks -- are gravity forces weaker in destinations with strong networks? To this end, we can parametrize $\delta(.)$ to include interactions between networks and distance terms:
\footnotesize
\begin{align}
    \delta(D_{ij},N_{ij}) &=  \delta_f 1\{D_{ij} \neq 0\} + \delta_v \log(D_{ij}) + \delta_{vs} \log(D_{ij})\times 1\{s(i) \neq s(j)\} \notag \\
    &+ \delta_{fn} 1\{D_{ij} \neq 0\}\times N_{ij} + \delta_{vn} \log(D_{ij})\times N_{ij} + \delta_{vsn} \log(D_{ij})\times 1\{s(i) \neq s(j)\}\times N_{ij} \label{eq:int},
\end{align}
\normalsize
where the $\delta_{.n}$ terms describe the additional moving cost that varies with destination networks. The ``net'' marginal effects of distance terms are thus $\delta_. + \delta_{.n}N_{ij}$, which depend on destination network size. In our preferred estimation, these $\delta_{.n}$ terms can be estimated by including additional instruments that interact our heat shock instrument with the distance terms.

\begin{figure}[H]
    \centering
        \caption{Effects of Distance by Destination Social Networks}
    \label{fig: mc}
    \includegraphics[width=\textwidth]{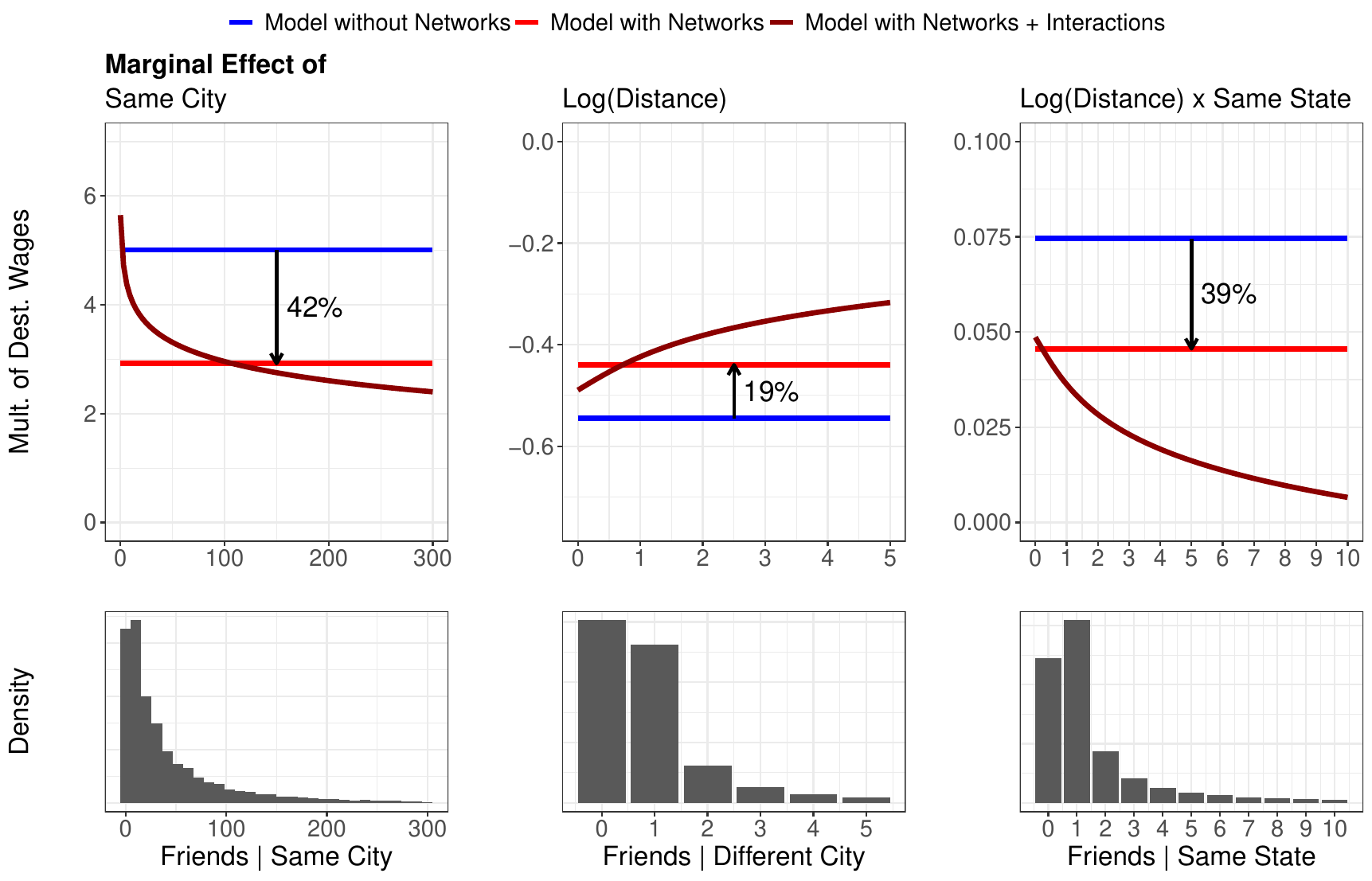}
    \begin{notes}
    This figure plots the net marginal effect of distance as a function of destination social networks, across distance terms (left, middle, right) and model specifications (blue, red, dark red). We first estimate our preferred specification (column 6 of Table \ref{tab: glm_compare_surv}) with interactions between distance and networks as in equation \ref{eq:int}. The dark red curve plots the estimated distance effects as a function of networks, scaled by the wage elasticity: $\delta(D_{ij},N_{ij})/\beta$. The blue and red curves plot distance coefficients from a regression of true moving costs $mc_{ij}=\gamma N_{ij} + \delta(D_{ij},N_{ij})$ on distance and network terms with no interactions (red) and distance terms only (blue) as in equations \ref{eq:mc1}-\ref{eq:mc2}. The bottom panels plot the distribution of social networks across origins (left), non-origins (middle), and in-state cities (right).
    \end{notes}
\end{figure}

Figure \ref{fig: mc} plots the net marginal effect of distance terms as a function of destination social networks. The dark red curves show that as destination networks increase, gravity forces weaken: the marginal effects of all distance terms (in Log wage units) get closer to zero ($\delta_{fn}<0,\delta_{vn}>0,\delta_{vsn}<0$). For example, in the left panel, increasing networks in the origin from 0 to 300 connections reduces the fixed cost of migrating outside the origin by 67\%, from roughly 6 times wages to 2 times wages. Increasing destination networks from 0 to 5 connections reduces the marginal cost of distance by 40\%, from $0.5$ times wages to $0.3$ times wages. This suggests that, in addition to providing level shifts in utility through $\gamma$, networks also reduce distance-related moving costs through $\delta_{.n}$.

How does accounting for networks change the gravity relationship? To this end, we first compute ``true'' moving costs $mc_{ij}$ from our preferred specification as the sum of all distance and network terms:
\begin{equation}\label{eq:mc1} mc_{ij} \equiv \gamma N_{ij} + \delta(D_{ij},N_{ij}). \end{equation}
Then, we compute alternative distance effects by regressing true moving costs $mc_{ij}$ on alternative specifications:
\begin{equation}\label{eq:mc2} mc_{ij} = \gamma^{NI} N_{ij} + \delta^{NI}(D_{ij}) + \epsilon^{NI}_{ij} \quad \text{or} \quad mc_{ij} = \delta^{NN}(D_{ij}) + \epsilon^{NN}_{ij},\end{equation}
where $\gamma^{NI}$ and $\delta^{NI}(.)$ capture network and distance effects under a model with no distance-network interactions; and $\delta^{NN}(.)$ capture distance effects under a model without networks entirely. By comparing $\delta(.)$ and $\delta^{NI}(.)$ to $\delta^{NN}(.)$ we can compute how gravity forces would change if we account for networks in the migration model.

In Figure \ref{fig: mc}, we find that accounting for networks dramatically reduces gravity forces, moving from a model without networks (blue) to a model with networks (red). In particular, the fixed cost of migration falls by 42\% (left), the marginal cost of distance falls by 19\% (middle), and the additional marginal cost of distance for out-of-state moves falls by 39\% (right). Taken together, these estimates suggest that a large portion of the gravity relationship can be explained by social networks. Previous estimates of gravity (that do not incorporate networks) may thus overestimate moving costs associated with physical distance.

\section{Spatial equilibrium}
\label{sec:ge}
Our model of migration above shows that social networks are an important determinant of migration choice, and explain a substantial portion of the gravity relationship in India. While this suggests expanding social networks may increase migration, the aggregate implications of this counterfactual are unclear. In addition to a labor supply response through migration, equilibrium forces such as labor demand responses, agglomeration, and congestion would impact the distribution of wages, amenities, and welfare. To study the effects of network counterfactuals, we thus develop a simple, static spatial equilibrium framework to account for these forces.\footnote{For tractability, we do not model dynamic spatial equilibrium.}

We borrow elements from prior work on tractable spatial equilibrium models (\cite{rr17}). Workers choose locations to maximize utility (governed by our migration model above). Firms hire workers to maximize profit, setting wages equal to the marginal product of labor, which is subject to agglomeration forces. Land markets determine local prices which depend on congestion forces. Like prices, local amenities also depend on amenity congestion forces. Finally, markets clear in equilibrium so that workers choosing locations to maximize utility, firms set wages and hire workers to maximizing profits, and land markets and amenities obey congestion.

\paragraph*{Environment}

Individuals $i$ select locations $j\in\mathcal J$ to maximize utility, subject to taste shocks and migration costs. Land is in fixed supply and is consumed by individuals. Firms in each location $j$ produce a numeraire, freely tradeable good with CRS technology subject and external economies of scale. 


\paragraph*{Preferences}

Individuals $i$ migrate to any $j$ after incurring migration cost $1/\tau_{ij}$. Preferences are Cobb-Douglas over housing $h_j$ and the consumption good $c_j$, whose prices are $r_j$ and $1$ respectively. Local amenities $\xi^A_j$ scale utility.:
\begin{equation}\label{eq:utility}
U_{ij} = \frac{1}{\tau_{ij}}\left( \frac{c_j}{1 - \psi} \right)^{1-\psi}\left( \frac{h_j}{\psi} \right)^{\psi}\nu_{ij}.
\end{equation}
Households inelastically supply one unit of labor, earn wage $w_j$, and spend a fixed portion of wages on housing under optimality:
\begin{equation}
\label{eq:hopt}
h_j = \frac{\psi w_j}{r_j}.
\end{equation}
Taste shocks $\nu_{ij}$ have a Frechet distribution with location parameter $\xi_j^A$ and scale parameter $\beta$. Thus, individual $i$ has indirect utility for location $j$ as:
\begin{equation}\label{eq:indirect_utility}
V_{ij} =\frac{1}{\tau_{ij}} \frac{w_j}{r_j^\psi}\nu_{ij}.
\end{equation}
Given the Frechet distribution, migration choice probabilities are written:
\begin{equation}\label{eq:choice_probabilities}
P_{ij}\equiv P(j = \text{argmax}_{j'} V_{ij}) = \frac{\xi_j^A(w_j/\tau_{ij})^\beta r_{j}^{-\psi\beta}}{\sum_k \xi_k^A(w_k/\tau_{ik})^\beta r_{k}^{-\psi\beta}}.
\end{equation}

\paragraph*{Firms}

Firms in each location $j$ produce a numeraire, freely tradeable good with CRS technology:
\begin{equation}\label{eq:production}
Y_j = Z_j L_j,
\end{equation}
where local productivity $Z_j$ is subject to external economies of scale $\phi$ depending on local labor supply $L_j$ (agglomeration forces):
\begin{equation}
    Z_j = \bar Z_j L_j^\phi.
\end{equation}
Profit maximizing firms set wages equal to their marginal product:
\begin{equation}\label{eq:wages}
w_j = Z_j.
\end{equation}

\paragraph*{Housing}

Each location has land $T_j$, which is fixed, and is used for housing.

\paragraph*{Amenities}

Amenities $\xi_j^A$ face congestion forces $\theta$ depending on local labor supply $L_j$:
\begin{equation}\label{eq:amen}
\xi_j^A = \kappa_j^\xi L_j^{-\theta}.
\end{equation}

\paragraph*{Equilibrium}

The spaital equilibrium is a vector of real wages $W_j = \frac{w_j}{r_j^\psi}$, populations $L_{j}$, and amenities $\xi_j^A$ such that:

\begin{enumerate}
    \item Households maximize utility as in \cref{eq:hopt} and \cref{eq:choice_probabilities}
    \item Firms maximize profits as in \cref{eq:wages}
    \item The land market clears:
    \begin{equation}\label{eq:land_market_clearance}
    T_j = h_j L_j \Rightarrow r_j = \frac{\psi w_j L_j}{T_j},
    \end{equation}
    given \cref{eq:hopt}.
    \item The labor market clears:\footnote{Immobile landlords in each location consume profits from land on the numeraire good.}
    \begin{equation}\label{eq:labor_market_clearance}
    L_j = \sum_i P_{ij}.
    \end{equation}
    \item Amenities obey congestion forces:
    \begin{equation}\label{eq:amen2}
\xi_j^A = \kappa_j^\xi L_j^{-\theta}.
\end{equation}
\end{enumerate}
All production is consumed exactly and the goods market clears by Walras' law.

\paragraph*{Characterization}

Denote the real wage $W_j$ as:
\begin{equation}\label{eq:real_wage}
W_j = \frac{w_j}{r_j^\psi} = \frac{w_j}{\left(\frac{\psi w_j L_j}{T_j}\right)^\psi}= (T_j/\psi)^\psi w_j^{1-\psi}L_j^{-\psi},
\end{equation}
where the second equality comes from land clearing given by \cref{eq:land_market_clearance}. Plugging in for wages under profit maximization given by \cref{eq:wages} this reduces to:
\begin{equation}\label{eq:kappaw}
W_j = \kappa_j^W L_j^{\phi(1-\psi)-\psi}.
\end{equation}
where $\kappa_j^W = (T_j/\psi)^\psi \bar Z^{1-\psi}$. Thus, choice probabilities can be characterized as:
\begin{equation}\label{eq:char_probabilities}
P_{ij} \propto \xi_j^A (W_j/\tau_{ij})^\beta,
\end{equation}
which, given the Frechet distribution, is equivalent to that of our log-additive migration model above given by \cref{eq:v}. The model's equilibrium is fully characterized by \cref{eq:labor_market_clearance}, \cref{eq:amen2}, \cref{eq:kappaw}, and \cref{eq:char_probabilities}, and can be solved by wage iteration. 






\subsection{Parametrization}
Table \ref{tab: ge_params} reports parameters used in our spatial equilibrium framework. Labor supply parameters that govern individual location decisions are estimated using our preferred specification of the migration model above. Due to data availability, we borrow equilibrium parameters from \cite{bm19}, which studies migration in a developing country (Indonesia) and itself borrows from prior estimates in the literature where unavailable.

\begin{table}[H]
    \centering
     \caption{Spatial Equilibrium Parameters}
     \label{tab: ge_params}
    \begin{tabular}{c c c}
    \toprule
    Parameter & Description & Value \\
    \midrule
    \multicolumn{3}{l}{\emph{Panel A: Labor Supply Parameters}} \\
       $\beta$  & Wage elasticity  & 1.11 \\
       $\xi_j^A$  & City amenities (initial)  & 0.0 -- 6.4  \\
       $\gamma$ & Network elasticity & 0.71   \\
       $\delta_f,\delta_v,\delta_{vs}$ & Distance elasticities  & 6.3,--0.55,0.05  \\
       $\delta_{fn},\delta_{vn},\delta_{vsn}$ & Distance $\times$ Network elasticities  & --0.57,0.08,--0.02 \\\\
    \multicolumn{3}{l}{\emph{Panel B: Equilibrium Parameters}} \\
       $\phi$ & Wage agglomeration & 0.05  \\
       $\psi$ & Price congestion & 0.30  \\
       $\theta$ & Amenity congestion & 0.02 \\
       \midrule\midrule\multicolumn{3}{l}{\emph{  }}
    \end{tabular}
    \begin{notes}
    This table reports parameter estimates used to specify the spatial equilibrium model. Panel A reports labor supply parameters estimated using our preferred specification of the migration model (column 6 of Table \ref{tab: glm_compare_surv}) with interactions between distance and networks as in equation \ref{eq:int}. Panel B reports equilibrium parameters borrowed from \cite{bm19}.
    \end{notes}
\end{table}

\subsection{Counterfactuals}
We wish to study the implications of social network expansion on equilibrium wages. Figure \ref{fig: counterfactuals} shows various changes to the geographic social network structure compared to an example baseline network and other counterfactuals (e.g. reducing distance or increasing wages). 

Panel A shows a baseline network for individual $i$ and potential connections to 14 cities, depicted by circles $j_1$-$j_{14}$. Some cities are connected (e.g. $j_2$) and some are unconnected (e.g. $j_7$). Some cities are closer in distance (e.g. $j_4$) and some cities are farther (e.g. $j_{14}$). The sizes of circles denote the number of social connections in the cities, and the darkness denotes their average wages.

Panel B depicts a scenario in which the variable cost of distance is zero (setting $\delta_v,\delta_{vn}=0$). This shrinks the effective distance from $i$ to each destination city $j$ to an infinitesimally small positive number (i.e. 0.0001KM). This is akin to an experiment that builds rapid transport infrastructure, as if any city $j$ was virtually zero kilometers away.

Panel C equalizes networks across cities to the same number of connections as the origin. This experiment increases $i$'s connections to all other cities, so that connections are equal across origin and destinations.

In contrast, Panel D fixes the total number of connections, but instead reallocates them to cities in the top 10\% of average wages. This is akin to an experiment in which all of $i$'s connections move to and are equally split between the richest 10\% of cities.

Panel E simply doubles the wages of cities in the top 10\% of average wages, simulating local productivity shocks. Finally, Panel F doubles both wages and networks in those cities, simulating local productivity and network shocks.

\begin{figure}[H]
    \centering
    \caption{Counterfactual Social Networks}
    \label{fig: counterfactuals}
    \begin{subfigure}[t]{0.45\textwidth}
    \caption{Baseline}
    \hspace{3em}
       \begin{tikzpicture}[scale=0.8]
    \Vertex[x=0,y=0,label=$i$,size=1.1,opacity=0.5]{A}
    \Vertex[x=-1,y=1,label=$j_4$,size=0.8,opacity=0.5]{B}
    \Vertex[x=-0.3,y=2,label=$j_2$,size=0.8]{C}
    \Vertex[x=1,y=1.2,label=$j_6$,opacity=0.2]{D}
    \Vertex[x=2.1,y=0.7,label=$j_9$,size=0.5,opacity=0.2]{DD}
    \Vertex[x=1.6,y=-1,label=$j_{12}$,opacity=0.2]{E}
    \Vertex[x=0.6,y=-1.2,label=$j_{10}$]{F}
    \Vertex[x=-1.3,y=-1.3,label=$j_{8}$,opacity=0.5]{G}
    \Vertex[x=-1.5,y=-0.3,label=$j_5$,size=0.5,opacity=0.2]{GD}
    \Vertex[x=-0.6,y=-2.1,label=$j_{14}$,size=0.8]{H}
    \Edge (A)(B)
    \Edge (A)(C)
    \Edge (A)(D)
    \Edge (A)(E)
    \Edge (A)(F)
    \Edge (A)(G)
    \Edge (A)(H)
    \Vertex[x=1.1,y=0.1,label=$j_7$,size=0.5,opacity=0.5]{I}
    \Vertex[x=1.1,y=-2,label=$j_{11}$,size=0.5,opacity=0.2]{J}
    \Vertex[x=-1.5,y=2,label=$j_1$,size=0.5]{K}
    \Vertex[x=1.25,y=2.1,label=$j_3$,size=0.5]{L}
\end{tikzpicture}
\end{subfigure}
\begin{subfigure}[t]{0.45\textwidth}
\caption{Zero Variable Distance Cost}
    \hspace{4em}
\begin{tikzpicture}[scale=0.8,baseline={(0,2)}]
    \Vertex[x=0,y=0,label=$i$,size=1.1,opacity=0.5]{A}
    \Vertex[x=-1/2,y=1/2,label=$j_4$,size=0.8,opacity=0.5]{B}
    \Vertex[x=-0.3/2,y=2/2,label=$j_2$,size=0.8]{C}
    \Vertex[x=1/2,y=1.2/2,label=$j_6$,opacity=0.2]{D}
    \Vertex[x=2.1/2,y=0.7/2,label=$j_9$,size=0.5,opacity=0.2]{DD}
    \Vertex[x=1.6/2,y=-1/2,label=$j_{12}$,opacity=0.2]{E}
    \Vertex[x=0.6/2,y=-1.2/2,label=$j_{10}$]{F}
    \Vertex[x=-1.3/2,y=-1.3/2,label=$j_{8}$,opacity=0.5]{G}
    \Vertex[x=-1.5/2,y=-0.3/2,label=$j_5$,size=0.5,opacity=0.2]{GD}
    \Vertex[x=-0.6/2,y=-2.1/2,label=$j_{14}$,size=0.8]{H}
    \Edge (A)(B)
    \Edge (A)(C)
    \Edge (A)(D)
    \Edge (A)(E)
    \Edge (A)(F)
    \Edge (A)(G)
    \Edge (A)(H)
    \Vertex[x=1.1/2,y=0.1/2,label=$j_7$,size=0.5,opacity=0.5]{I}
    \Vertex[x=1.1/2,y=-2/2,label=$j_{11}$,size=0.5,opacity=0.2]{J}
    \Vertex[x=-1.5/2,y=2/2,label=$j_1$,size=0.5]{K}
    \Vertex[x=1.25/2,y=2.1/2,label=$j_3$,size=0.5]{L}
\end{tikzpicture}
\end{subfigure}
\begin{subfigure}[t]{0.45\textwidth}
    \caption{Equal Networks}
    \hspace{2.5em}
       \begin{tikzpicture}[scale=0.8]
    \Vertex[x=0,y=0,label=$i$,size=1.1,opacity=0.5]{A}
    \Vertex[x=-1,y=1,label=$j_4$,size=1.1,opacity=0.5]{B}
    \Vertex[x=-0.3,y=2,label=$j_2$,size=1.1]{C}
    \Vertex[x=1,y=1.2,label=$j_6$,opacity=0.2,size=1.1]{D}
    \Vertex[x=2.1,y=0.7,label=$j_9$,size=1.1,opacity=0.2]{DD}
    \Vertex[x=1.6,y=-1,label=$j_{12}$,opacity=0.2,size=1.1]{E}
    \Vertex[x=0.6,y=-1.2,label=$j_{10}$,size=1.1]{F}
    \Vertex[x=-1.3,y=-1.3,label=$j_{8}$,opacity=0.5,size=1.1]{G}
    \Vertex[x=-1.5,y=-0.3,label=$j_5$,size=1.1,opacity=0.2]{GD}
    \Vertex[x=-0.6,y=-2.1,label=$j_{14}$,size=1.1]{H}
    \Edge (A)(B)
    \Edge (A)(C)
    \Edge (A)(D)
    \Edge (A)(E)
    \Edge (A)(F)
    \Edge (A)(G)
    \Edge (A)(H)
    \Vertex[x=1.1,y=0.1,label=$j_7$,size=1.1,opacity=0.5]{I}
    \Vertex[x=1.1,y=-2,label=$j_{11}$,size=1.1,opacity=0.2]{J}
    \Vertex[x=-1.5,y=2,label=$j_1$,size=1.1]{K}
    \Vertex[x=1.25,y=2.1,label=$j_3$,size=1.1]{L}
    \Edge (A)(I)
    \Edge (A)(J)
    \Edge (A)(K)
    \Edge (A)(L)
    \Edge (A)(DD)
    \Edge (A)(GD)
\end{tikzpicture}
\end{subfigure}
\begin{subfigure}[t]{0.45\textwidth}
\caption{Network Reallocation}
    \hspace{2.5em}
\begin{tikzpicture}[scale=0.8]
    \Vertex[x=0,y=0,label=$i$,size=.5,opacity=0.5]{A}
    \Vertex[x=-1,y=1,label=$j_4$,size=.5,opacity=0.5]{B}
    \Vertex[x=-0.3,y=2,label=$j_2$,size=0.8]{C}
    \Vertex[x=1,y=1.2,label=$j_6$,opacity=0.2,size=.5]{D}
    \Vertex[x=2.1,y=0.7,label=$j_9$,size=0.5,opacity=0.2]{DD}
    \Vertex[x=1.6,y=-1,label=$j_{12}$,opacity=0.2,size=.5]{E}
    \Vertex[x=0.6,y=-1.2,label=$j_{10}$,size=0.8]{F}
    \Vertex[x=-1.3,y=-1.3,label=$j_{8}$,opacity=0.5,size=.5]{G}
    \Vertex[x=-1.5,y=-0.3,label=$j_5$,size=0.5,opacity=0.2]{GD}
    \Vertex[x=-0.6,y=-2.1,label=$j_{14}$,size=0.8]{H}
    \Vertex[x=1.1,y=0.1,label=$j_7$,size=0.5,opacity=0.5]{I}
    \Vertex[x=1.1,y=-2,label=$j_{11}$,size=0.5,opacity=0.2]{J}
    \Vertex[x=-1.5,y=2,label=$j_1$,size=0.8]{K}
    \Vertex[x=1.25,y=2.1,label=$j_3$,size=0.8]{L}
    \Edge (A)(K)
    \Edge (A)(C)
    \Edge (A)(L)
    \Edge (A)(F)
    \Edge (A)(H)
\end{tikzpicture}
\end{subfigure}
\begin{subfigure}[t]{0.45\textwidth}
    \caption{Double Top Wages}
    \hspace{3em}
 \begin{tikzpicture}[scale=0.8]
    \Vertex[x=0,y=0,label=$i$,size=1.1,opacity=0.5]{A}
    \Vertex[x=-1,y=1,label=$j_4$,size=0.8,opacity=0.5]{B}
    \Vertex[x=-0.3,y=2,label=$j_2$,size=0.8,color={rgb:red,1;green,2;blue,3},opacity=0.8]{C}
    \Vertex[x=1,y=1.2,label=$j_6$,opacity=0.2]{D}
    \Vertex[x=2.1,y=0.7,label=$j_9$,size=0.5,opacity=0.2]{DD}
    \Vertex[x=1.6,y=-1,label=$j_{12}$,opacity=0.2]{E}
    \Vertex[x=0.6,y=-1.2,label=$j_{10}$,color={rgb:red,1;green,2;blue,3},opacity=0.8]{F}
    \Vertex[x=-1.3,y=-1.3,label=$j_{8}$,opacity=0.5]{G}
    \Vertex[x=-1.5,y=-0.3,label=$j_5$,size=0.5,opacity=0.2]{GD}
    \Vertex[x=-0.6,y=-2.1,label=$j_{14}$,size=0.8,color={rgb:red,1;green,2;blue,3},opacity=0.8]{H}
    \Edge (A)(B)
    \Edge (A)(C)
    \Edge (A)(D)
    \Edge (A)(E)
    \Edge (A)(F)
    \Edge (A)(G)
    \Edge (A)(H)
    \Vertex[x=1.1,y=0.1,label=$j_7$,size=0.5,opacity=0.5]{I}
    \Vertex[x=1.1,y=-2,label=$j_{11}$,size=0.5,opacity=0.2]{J}
    \Vertex[x=-1.5,y=2,label=$j_1$,size=0.5,color={rgb:red,1;green,2;blue,3},opacity=0.8]{K}
    \Vertex[x=1.25,y=2.1,label=$j_3$,size=0.5,color={rgb:red,1;green,2;blue,3},opacity=0.8]{L}
\end{tikzpicture}
\end{subfigure}
\begin{subfigure}[t]{0.45\textwidth}
\caption{Double Top Wages and Networks}
    \hspace{3em}
\begin{tikzpicture}[scale=0.8]
    \Vertex[x=0,y=0,label=$i$,size=1.1,opacity=0.5]{A}
    \Vertex[x=-1,y=1,label=$j_4$,size=0.8,opacity=0.5]{B}
    \Vertex[x=-0.3,y=2,label=$j_2$,size=1.1,color={rgb:red,1;green,2;blue,3},opacity=0.8]{C}
    \Vertex[x=1,y=1.2,label=$j_6$,opacity=0.2]{D}
    \Vertex[x=2.1,y=0.7,label=$j_9$,size=0.5,opacity=0.2]{DD}
    \Vertex[x=1.6,y=-1,label=$j_{12}$,opacity=0.2]{E}
    \Vertex[x=0.6,y=-1.2,label=$j_{10}$,color={rgb:red,1;green,2;blue,3},opacity=0.8,size=0.8]{F}
    \Vertex[x=-1.3,y=-1.3,label=$j_{8}$,opacity=0.5]{G}
    \Vertex[x=-1.5,y=-0.3,label=$j_5$,size=0.5,opacity=0.2]{GD}
    \Vertex[x=-0.6,y=-2.1,label=$j_{14}$,size=1.1,color={rgb:red,1;green,2;blue,3},opacity=0.8]{H}
    \Edge (A)(B)
    \Edge (A)(C)
    \Edge (A)(D)
    \Edge (A)(E)
    \Edge (A)(F)
    \Edge (A)(G)
    \Edge (A)(H)
    \Vertex[x=1.1,y=0.1,label=$j_7$,size=0.5,opacity=0.5]{I}
    \Vertex[x=1.1,y=-2,label=$j_{11}$,size=0.5,opacity=0.2]{J}
    \Vertex[x=-1.5,y=2,label=$j_1$,color={rgb:red,1;green,2;blue,3},opacity=0.8]{K}
    \Vertex[x=1.25,y=2.1,label=$j_3$,color={rgb:red,1;green,2;blue,3},opacity=0.8]{L}
    \Edge (A)(K)
    \Edge (A)(L)
\end{tikzpicture}
\vspace{0.5em}
\end{subfigure}
\begin{notes}
This figure depicts visual intuition for network counterfactuals. Each panel an example of a modified individual networks across cities. Panel A shows an example network at baseline, while Panels (B-F) show modified networks. Each bubble is a city $j_1,\dots,j_{12}$. The central bubble denoted by $i$ is individual $i$'s origin city. Lines indicate whether or not individual $i$ is connected to the respective city. The size of the bubble denote the size of $i$'s network in the given city. Darker (lighter) bubbles denote cities with higher (lower) average wages. In Panel B, we set the variable cost of distance to zero ($\delta_v,\delta_{vn}=0$). In Panel C, we set network sizes in all cities equal to that of the origin. In Panel D, assign all friends equally to cities in the top 10\% of average wages. In Panel E, we double the wages of cities in the top 10\% of average wages. In Panel F, we both double network size and wages of cities in the top 10\% of average wages.
\end{notes}
\end{figure}

These counterfactuals are shocks to initial values of $\tau_{ij}$ or $W_j$, which results in a new equilibrium vector of population, wages, and amenities $\{L_j,W_j,\xi_j^A\}$ that satisfy \cref{eq:labor_market_clearance}, \cref{eq:amen2}, \cref{eq:kappaw}, and \cref{eq:char_probabilities}. These are computed by iteratively updating each variable until convergence. We can compare these new equilibria against baseline (initial values from observed data) to study the aggregate impacts of these counterfactuals.

Table \ref{tab: effect_cf} reports the impacts of these counterfactual scenarios on equilibrium outcomes relative to the baseline (Panel A). In Panel B, we find that eliminating the variable costs associated with distance (akin to improving transport infrastructure) increases rates of migration by 2.6 times and distance from origin by 5.7 times. This increase in migration results in a 3\% increase in average wages, a 10\% increase in amenities, and a 9\% increase in welfare. We find that wage gains are concentrated among individuals who live in cities in the bottom-quartile of the average wage distribution, with a 17\% increase in wages and 12\% increase in welfare. By construction, individuals who live in cities at the top-quartile \emph{ex-ante} have the lowest to gain in terms of wages from reducing moving costs.

In Panel C, we test the counterfactual of equalizing networks across cities, such that individuals have the same number of connections in each destination city as they do in their origin city. This increases the total number of connections by 20 times, given how few connections individuals have outside their origin city, which increases migration rates by 4.7 times and distances by 12 times. This network-driven increase in migration is due to two forces: (1) destinations becoming relatively more attractive due to main network effects via $\gamma$; and (2) distance-related moving costs become smaller via $\delta_{.n}$, which makes destinations effectively ``closer''. This increase in migration results in a 3\% increase in average wages, a 22\% increase in amenities, and a 21\% increase in welfare. For the bottom-quartile wage group, average wages increase by 24\%. In Panel D, we find that the network reallocation counterfactual delivers large returns as well, while keeping total networks fixed for each individual. Relative to the zero variable distance cost counterfactual, equalizing or reallocating networks across cities delivers even larger gains, especially so for those in the bottom-quartile.

Finally, in Panel E we see that doubling the average wages of top-quartile cities (akin to local productivity shocks) increases wages by 52\% and welfare by 22\% overall. However, gains are concentrated among those who live in top-quartile cities \emph{ex-ante} as opposed to new migrants from bottom-quartile cities. In Panel F, we find that doubling networks as well as wages for the top-quartile cities leads to twice the wage gain for bottom-quartile cities.

\newpage

\begin{landscape}
\begin{table}[htbp]
\scriptsize
\vspace{-2em}
\caption{Equilibrium Impacts of Counterfactuals}
\label{tab: effect_cf}
\begin{tabular}{lccccccc}
\toprule
\multicolumn{2}{l}{\emph{Panel A: Baseline}} & \multicolumn{6}{c}{Level at Baseline} \\
 & Network Size & Migration Rate & Distance From Origin (KM) & Mean Wage (USD) & SD Wage (USD) & Amenities (\%Wages) & Total Welfare (\%Wages) \\ 
  \hline
All & 83.24 & 0.04 & 8.60 & 9328.19 & 4199.10 & -10.38 & 6.11 \\ 
  Bottom Quartile Wage & 67.57 & 0.05 & 18.56 & 4226.46 & 1963.98 & -10.24 & 5.41 \\ 
  Top Quartile Wage & 101.81 & 0.03 & 5.09 & 14831.11 & 1203.04 & -10.28 & 6.84 \\ 
   \hline
\\
\multicolumn{2}{l}{\emph{Panel B: Zero Variable Distance Cost}} & \multicolumn{6}{c}{Multiple of Baseline} \\
 & Network Size & Migration Rate & Distance From Origin & Mean Wage & SD Wage & Amenities & Total Welfare \\ 
  \hline
All & 1.00 & 2.56 & 5.74 & 1.03 & 0.94 & 1.10 & 1.09 \\ 
  Bottom Quartile Wage & 1.00 & 2.92 & 4.42 & 1.17 & 1.05 & 1.10 & 1.12 \\ 
  Top Quartile Wage & 1.00 & 1.94 & 4.45 & 1.00 & 0.99 & 1.05 & 1.06 \\ 
   \hline
\\
\multicolumn{2}{l}{\emph{Panel C: Equal Networks}} & \multicolumn{6}{c}{Multiple of Baseline} \\
 & Network Size & Migration Rate & Distance From Origin & Mean Wage & SD Wage & Amenities & Total Welfare \\ 
  \hline
All & 21.47 & 4.67 & 11.99 & 1.03 & 0.90 & 1.22 & 1.21 \\ 
  Bottom Quartile Wage & 18.47 & 3.98 & 6.61 & 1.24 & 1.20 & 1.14 & 1.21 \\ 
  Top Quartile Wage & 23.25 & 5.28 & 15.95 & 0.98 & 1.06 & 1.18 & 1.20 \\ 
   \hline
\\
\multicolumn{2}{l}{\emph{Panel D: Network Reallocation}} & \multicolumn{6}{c}{Multiple of Baseline} \\
 & Network Size & Migration Rate & Distance From Origin & Mean Wage & SD Wage & Amenities & Total Welfare \\ 
  \hline
All & 1.00 & 2.29 & 7.97 & 1.07 & 0.98 & 1.01 & 0.66 \\ 
  Bottom Quartile Wage & 1.00 & 2.51 & 5.23 & 1.28 & 1.45 & 0.98 & 0.64 \\ 
  Top Quartile Wage & 1.00 & 1.84 & 6.47 & 1.03 & 1.16 & 1.04 & 0.80 \\ 
   \hline
\\
\multicolumn{2}{l}{\emph{Panel E: Double Top Wages}} & \multicolumn{6}{c}{Multiple of Baseline} \\
 & Network Size & Migration Rate & Distance From Origin & Mean Wage & SD Wage & Amenities & Total Welfare \\ 
  \hline
All & 1.00 & 0.99 & 1.27 & 1.52 & 2.49 & 1.00 & 1.22 \\ 
  Bottom Quartile Wage & 1.00 & 1.10 & 1.22 & 1.11 & 1.79 & 1.00 & 1.01 \\ 
  Top Quartile Wage & 1.00 & 1.00 & 0.87 & 2.00 & 2.06 & 1.00 & 1.69 \\ 
   \hline
\\
\multicolumn{2}{l}{\emph{Panel F: Double Top Wages and Networks}} & \multicolumn{6}{c}{Multiple of Baseline} \\
 & Network Size & Migration Rate & Distance From Origin & Mean Wage & SD Wage & Amenities & Total Welfare \\ 
  \hline
All & 1.42 & 1.25 & 1.86 & 1.54 & 2.47 & 1.01 & 1.27 \\ 
  Bottom Quartile Wage & 1.17 & 1.41 & 1.63 & 1.21 & 1.93 & 1.02 & 1.04 \\ 
  Top Quartile Wage & 1.87 & 1.15 & 1.09 & 2.01 & 2.03 & 1.01 & 1.79 \\ 
   \hline
 \vspace{-1.2em}\\
\midrule
\end{tabular}
\begin{notes}
This table reports estimates of the impacts of various counterfactuals on equilibrium outcomes from the parametrized spatial equilibrium model. Panel A reports baseline levels of each outcome, split by income group: all individuals, those who live in a city in the bottom quartile of the average wage distribution across cities, and those who live in the top quartile. Panels B-F report equilibrium outcomes from various counterfactuals described in Figure \ref{fig: counterfactuals}. For these panels, outcomes are reported as a multiple of the baseline level for comparison. For each scenario, we find values of variables that satisfy \cref{eq:labor_market_clearance}, \cref{eq:amen2}, \cref{eq:kappaw}, and \cref{eq:char_probabilities} using model parameters in Table \ref{tab: ge_params}. We simulate draws of $\epsilon_{ij}$ in the location choice and keep those that satisfy utility maximization of individuals. Estimates are reported as averages over the empirical distribution of kept draws. Estimates come from a 0.1\% sample of users.
\end{notes}
\end{table}
\end{landscape}

\section{Mechanisms}
\label{sec:mech}
Our models and data suggest social networks are an important determinant of migration, and that expansions to these networks may deliver aggregate economic returns. The mechanisms behind these network effects themselves are unclear and may reflect multiple forces (\cite{b89}). For example, informal networks provide economic support that may correct for market imperfections commonly seen in developing countries (e.g. credit, insurance, information, housing). Alternatively, social connections may provide emotional support that simply reflect individual taste (e.g. leisure, socialization). The implications may be policy relevant. If social support is entirely economic, governments may consider expanding social insurance, credit markets, or reducing job-search frictions in order to correct market imperfections that induce the reliance on informal networks. If it is emotional and thus taste-based, governments may consider investing in community-building programs and other policies that may expand social networks themselves.

While we do not observe these mechanisms directly, we attempt to study them in two ways: first, we study heterogeneity in networks effects along economic characteristics at the destination -- that is, if network effects are smaller when destinations have higher wages, migrant-friendly policies, or better access to formal credit; and second, we administer a qualitative survey among 50,000 Facebook users to understand which components of social support are the most important to individuals.

Lastly, in contrast to studying what drives the \emph{effects} of networks, we study what drives the \emph{levels} of social networks themselves. We exploit unique data on evolution of social networks over the life cycle to provide suggestive evidence of the effects of college attendance on social capital outcomes.

\subsection{Heterogeneity by destination economic characteristics}
We augment our migration model by including observed heterogeneity in the network effect $\gamma$ that varies by destination characteristics $X_{kj}$:
\begin{equation} \gamma_j = \gamma_0 + \sum_k \gamma_k X_{kj}.\end{equation}
Given limited data in India, we focus on three destination characteristics that capture potential sources of economic support for migrants: the average wage, the number of bank branches, and an index for migrant-friendly policy. Wages capture how higher earnings can provide a vector of economic support for individuals. Bank branches capture formal credit access, which may reduce the reliance on informal lending through social networks. Finally, migrant-friendly policy captures how local government policy may reduce a host of market frictions (e.g. information, job search, or housing) that networks provide. If $\gamma_k<0$, then improvements in these destination characteristics may reduce the network-migration relationship, suggestive of an economic support mechanism in preferences for social networks.

As above, data on average wages by destination come from the Periodic Labour Force Survey (PLFS). Data on bank branches from the 2017 Reserve Bank of India (RBI) database, geocoded to the district level. The migrant-friendly policy index is constructed by assessing state-level policies across government sectors as in \cite{asss20}.\footnote{The index is a composite measure across sector-specific indexes including children's rights, education, health, housing, identity and registration, labor market integration, political participation, and social benefits.}

To address network endogeneity in the interaction terms, we estimate parameters $\gamma_k$ by including additional interacted instruments as we do for estimating the distance interactions $\delta_{.n}$. However, we note that characteristics $X_{kj}$ are not exogenous, so we interpret estimates as ``conditional'' network effects.

Table \ref{tab: dhet} reports estimates of network effects by destination characteristics. We find that improvements to economic outcomes of the destination lead to statistically-significant reductions in the network effect. The negative signs are consistent with an economic mechanism of social support, where if formal economic conditions improve, there is less reliance on informal social networks to supplement them. Magnitudes are economically meaningful: at the mean of other variables, moving from the 25th to the 75th percentile of wages, bank branches, and migrant-friendly policy reduces the network effect by 6\%, 11\%, and 26\%, respectively. 

\begin{table}[H]
\centering
\caption{Network Effects by Destination Characteristics}
\label{tab: dhet}
\begin{tabular}{lc}
\toprule
Dependent Variable:&Choice$_t$\\
  & Heat IV\\
Model:&(1)\\
\midrule Log(Dest. Friends)$_{t-1}$ & 2.8$^{***}$\\
  &(0.39)\\
Log(Dest. Friends)$_{t-1}$ $\times $ Log(Dest. Wage) & --0.03$^{*}$\\
  &(0.02)\\
Log(Dest. Friends)$_{t-1}$ $\times $ Log(Dest. Bank Branches) & --0.07$^{***}$\\
  &(0.01)\\
Log(Dest. Friends)$_{t-1}$ $\times $ Log(Dest. Migrant-Friendly Index) & --0.44$^{***}$\\
  &(0.09)\\
\midrule Distance Controls & Y\\
Distance $\times$ Network Controls & Y\\
\midrule \emph{Fixed-effects}&  \\
Indiv. $\times$ Year & Y\\
Dest. & Y\\
\midrule Observations & 4M\\
\midrule\midrule\multicolumn{2}{l}{\emph{  }}\\
\end{tabular}
\begin{notes}
This table reports estimates of model coefficients using our preferred specification of the migration model (column 6 of Table \ref{tab: glm_compare_surv}) with interactions between distance and networks and between networks and three destination characteristics. Government data on wages and bank branches by district come from the 2017-18 round of the Periodic Labour Force Survey and the 2017 Reserve Bank of India database, respectively. The migrant-friendly index by state comes from \cite{asss20}. Standard errors are in paranthesis and clustered at the individual-by-year level.\\
*$p<0.10$, **$p<0.05$, ***$p<0.01$
\end{notes}
\end{table}

\paragraph{Natural experiment: the expansion of bank branches}

Because destination characteristics $X_{kj}$ are not randomly assigned, these effects may capture other components of destinations that are correlated with $X_{kj}$, potentially biasing estimates of interaction terms. To assess, this we exploit a natural experiment in India that expanded credit access in some districts and not in others. In particular, India implemented the 2005 Banking Regulation Act, which incentivized the private sector to open bank branches in ``unbanked'' districts and not in ``banked'' districts. This categorization was decided using a cutoff based on the population-per-branch in each district in 2005, which resulted in a discontinuous increase in bank branches a decade later in ``unbanked'' districts above this cutoff. This policy variation has been used in prior research to study the impacts of formal credit access in India (\cite{y17}; \cite{km20}).

We can exploit this variation using a similar regression discontinuity design to study the effects of formal credit expansion on the network effect of migration. In particular, we can interact the network effect $\gamma$ with whether or not a district was marginally ``unbanked'' -- just above vs. just below the cutoff. In particular, we include network interactions with the running variable (population-per-branch in 2005) and an indicator for being above the cutoff (unbanked in 2005), restricting to districts within a narrow bandwidth of the cutoff:
\begin{equation} \gamma_j = \gamma_0 + \gamma_1 \text{Population-per-branch}_j + \gamma_2 \text{Unbanked}_j.\end{equation}
The term $\gamma_2$ captures the additional network effect of being above the cutoff (unbanked) and therefore being exposed to a greater expansion of formal credit.

Table \ref{tab: bank_est} reports estimates from this natural experiment. Column (1) reports the first-stage: we find that in a narrow interval around the cutoff, there is a statistically significant increase in private bank branches in 2017 (roughly 25 more branches) for districts that are marginally unbanked in 2005. Column (2) reports the reduced-form: we find that being marginally unbanked (and thus an increase in the number of private bank branches) results in negative but statistically insignificant impact to the network effect of migration. While we cannot reject a zero effect size, scaling this point estimate by the first-stage effect suggests moving from the 25th to the 75th percentile of private banks reduces the network effect by 18\%.

Taken together, our estimates of heterogeneity suggest economic improvements in the destination may reduce the reliance on informal social networks in migration choice. This suggests the importance of economic support in network effects, and the potential role for governments to reduce network-related moving costs through economic growth, formal credit access, and migrant-friendly policy.

\begin{table}[H]
\centering
\caption{Network Effects and Banking Expansion}
\label{tab: bank_est}
\begin{tabular}{lcc}
\toprule
Dependent Variables:& Dest. Pvt. Branches&Choice$_t$\\
  &   & Heat IV\\
Model:&(1) & (2)\\
 &  OLS  & Logit\\
\midrule Log(Dest. Friends)$_{t-1}$ &     & --0.30\\
  &    & (0.47)\\
Log(Dest. Friends)$_{t-1}$ $\times $ Dest. Pop. Per Branch &    & 0.08$^{**}$\\
  &    & (0.04)\\
Log(Dest. Friends)$_{t-1}$ $\times $ 1$\{$Dest. Unbanked$\}$ &    & --0.08\\
  &   & (0.12)\\
Dest. Pop. Per Branch & --12.2$^{***}$   &   \\
  &(3.7)   &   \\
  1$\{$Dest. Unbanked$\}$ & 24.7$^{*}$  &   \\
  &(14.2)   &   \\
\midrule Bandwidth & 3.3 & 3.3\\
\midrule Distance Controls &   & Y\\
Distance $\times$ Network Controls &   & Y\\
\midrule \emph{Fixed-effects}&      &  \\
Indiv. $\times$ Year &   & Y\\
Dest. &   & Y\\
\midrule Observations & 185&1M\\
\midrule\midrule\multicolumn{3}{l}{\emph{  }}\\
\end{tabular}
\begin{notes}
This table reports estimates of the effects of India's 2005 Bank Regulation Act on private bank branches and network effects in migration choice. Column (1) reports estimates from an OLS regression at the district-level of private bank branches in 2017 on the running variable (population-per-branch in 2005) and a treatment indicator for being above the cutoff (unbanked in 2005). Column (2) reports estimates of network effects interacted with these terms, using our preferred specification of the migration model (column 6 of Table \ref{tab: glm_compare_surv}) with interactions between distance and networks and between networks and destination banking variables in Column 1. Estimates are restricted to districts within a bandwidth of 3.3 persons-per-branch above and below the cutoff. Bank branch data come from the Reserve Bank of India database. Standard errors are in parenthesis and clustered at the district level.\\
*$p<0.10$, **$p<0.05$, ***$p<0.01$
\end{notes}
\end{table}

\subsection{Qualitative survey responses}
\begin{figure}[H]
    \centering
    \caption{Survey Evidence on Social Support}
    \label{fig: surv_support}
    \includegraphics[width=\textwidth]{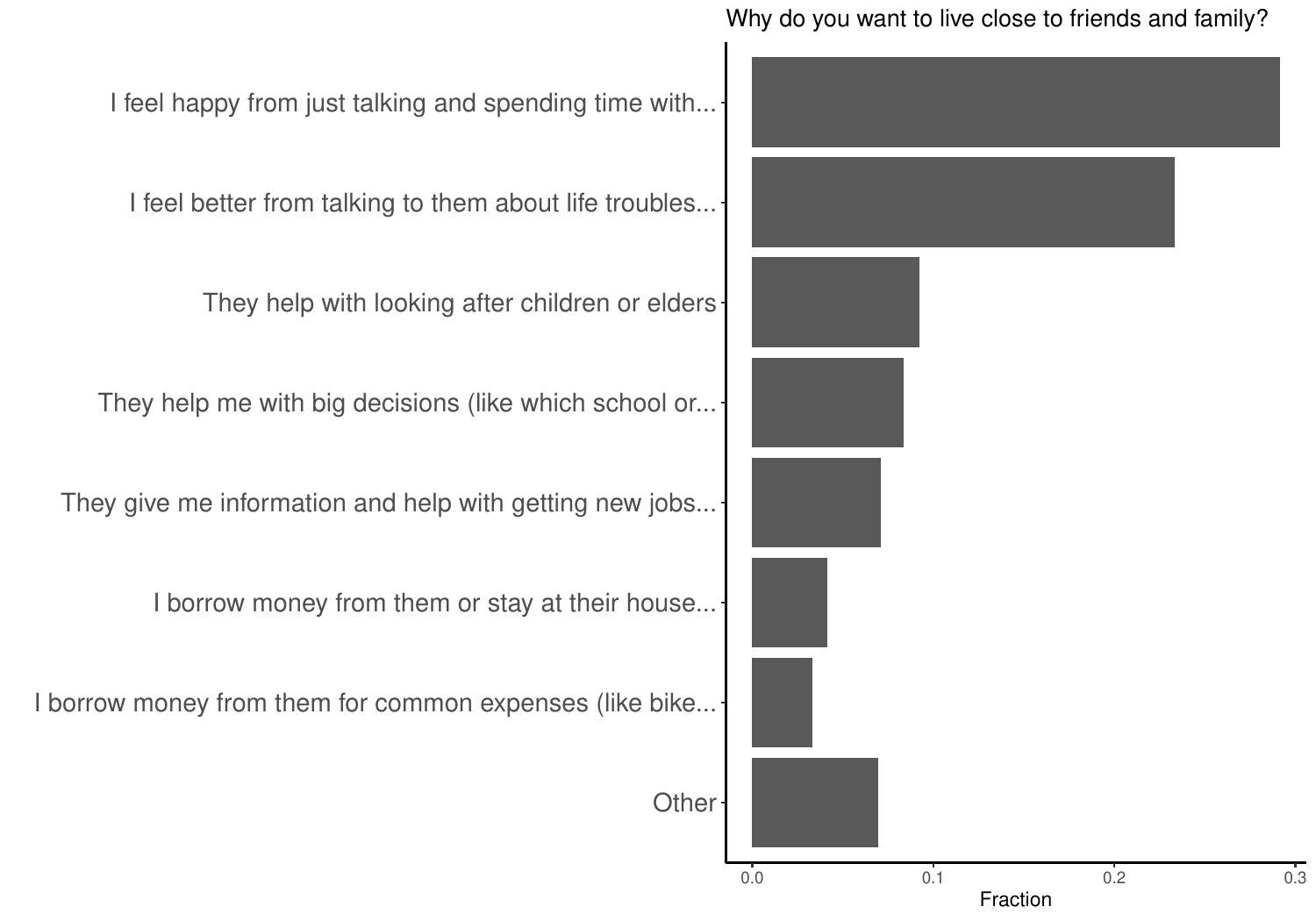}
    \begin{notes}
    This figure reports survey frequencies across categories for the question ``what is the most important reason you want to live close to friends and family?'' The survey was administered among a random sample of 50,000 Indian Facebook users in 2022.
    \end{notes}
\end{figure}

While they are significant, economic characteristics cannot explain the entire heterogeneity in network effects across destinations. To complement this, we administered a survey to capture beliefs regarding social support. Figure \ref{fig: surv_support} reports the distribution of responses across categories to the following question:
\begin{quote}
    \emph{What is the most important reason you want to live close to friends/family?}
\end{quote}
As their most important source of support, we find that 52\% of respondents list emotional support such as happiness from spending time together or talking during stressful times; and 32\% of respondents list economic support such as information or assistance (with child/elderly care, education, employment, or marriage opportunities), informal loans (for common expenses or during emergencies), or access to housing during unexpected shocks. This suggests an important role for emotional support -- in addition to economic support -- in driving preferences for social networks. 

\subsection{Education and social capital formation}
Our survey evidence suggests emotional support may be important, consistent with persistent networks effects even for destinations that are economically advanced (high wages, access to formal banks, and migrant-friendly policy). One conclusion may be that, due to the emotional support provided by social networks, economic policy alone may not fully eliminate the reliance on informal networks in migration choice. Given these strong preferences for social networks, as well as the economic returns to network expansion, this begs the question: what policies can expand social networks?

In our descriptive findings, we show that individuals that attended college have larger and richer networks compared to those that did not. Indeed, much work suggests education is an important dimension of social capital formation. However, the effects of education access on network formation remain largely unquantified, especially so in developing countries where informal social networks may be more important.\footnote{An exception is \cite{osa20}, which uses similar data from Facebook to study network formation among U.S. college students.} 

To quantify these effects, we compare the growth of social capital outcomes between college-attendees and non-college-attendees throughout their life-cycle. We estimate the effects of college attendance using a ``panel difference-in-differences'' specification for individual $i$ with age $a$ in year $t$:
\begin{equation}\label{eq:did_ed}Y_{iat} = \sum_{a'=15}^{30} \beta_{a'} 1\{\text{College Attended}\}_i\times 1\{a=a'\} + \eta_i + \delta_a + \omega_t + \epsilon_{iat},\end{equation}
where $Y_{iat}$ is an individual outcome of interest and $1\{\text{College Attended}\}_i$ is whether individual $i$ ever attended college. The parameter $\beta_a$ captures the additional effect of college attendance on $Y$ at age $a$. We expect that these additional effects should be close to zero before age 17, the typical age of college attendance in India, and increase thereafter.

\begin{figure}[H]
    \centering
    \caption{Social Capital and College Attendance}
    \label{fig: dd_col}
    \includegraphics[width=0.55\textwidth]{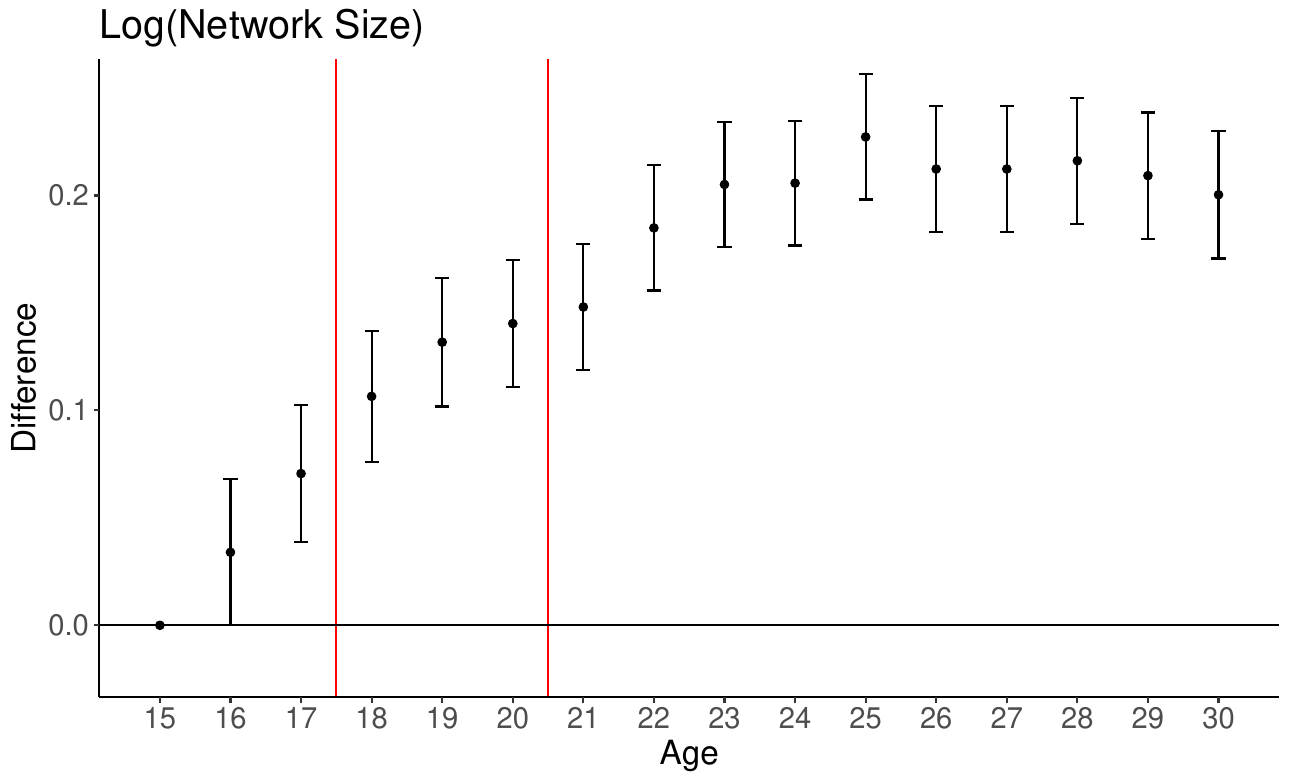}
    \includegraphics[width=0.55\textwidth]{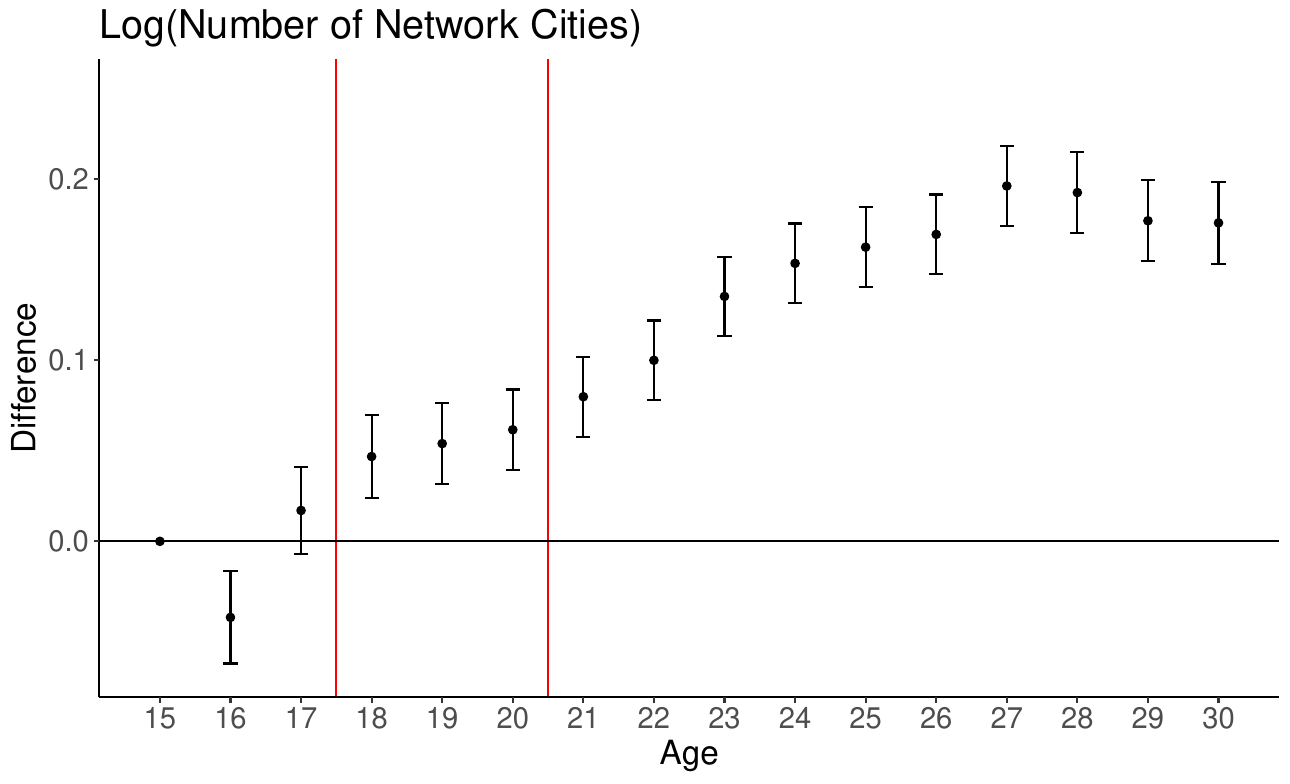}
    \includegraphics[width=0.55\textwidth]{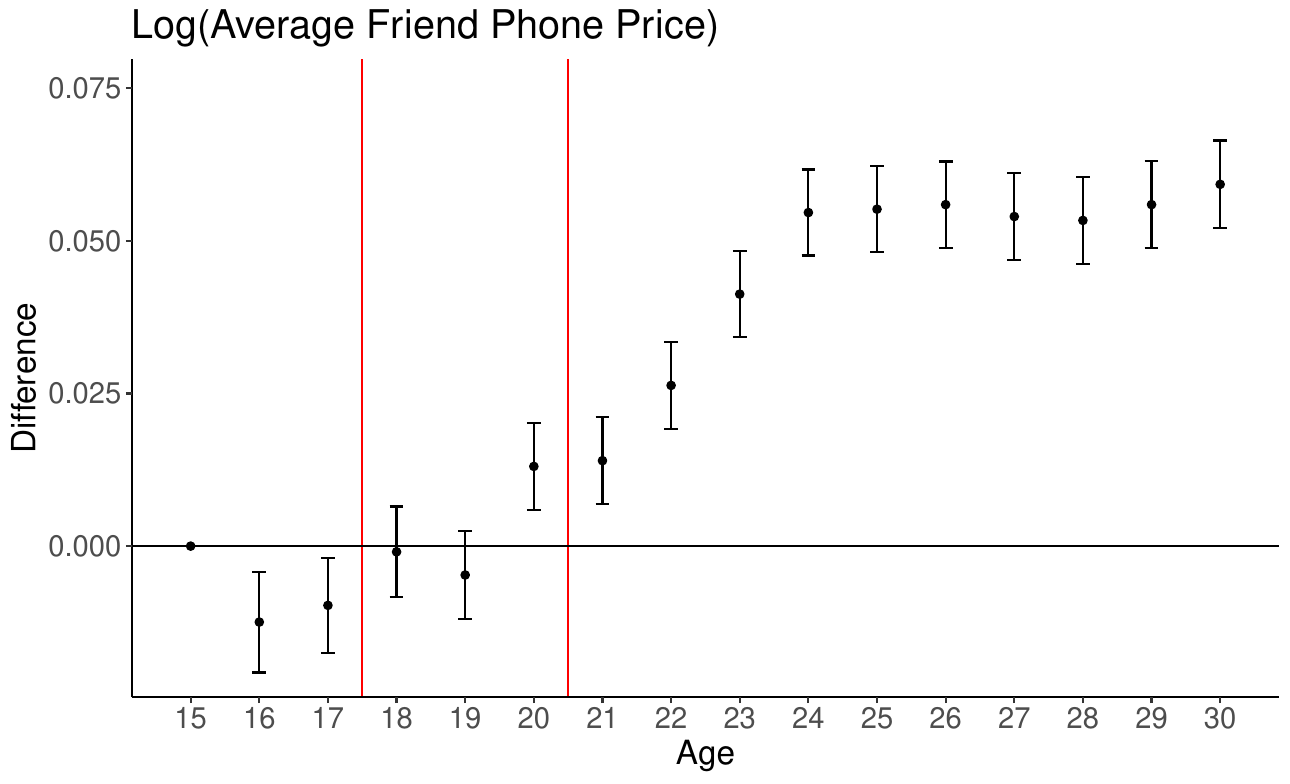}
    \includegraphics[width=0.55\textwidth]{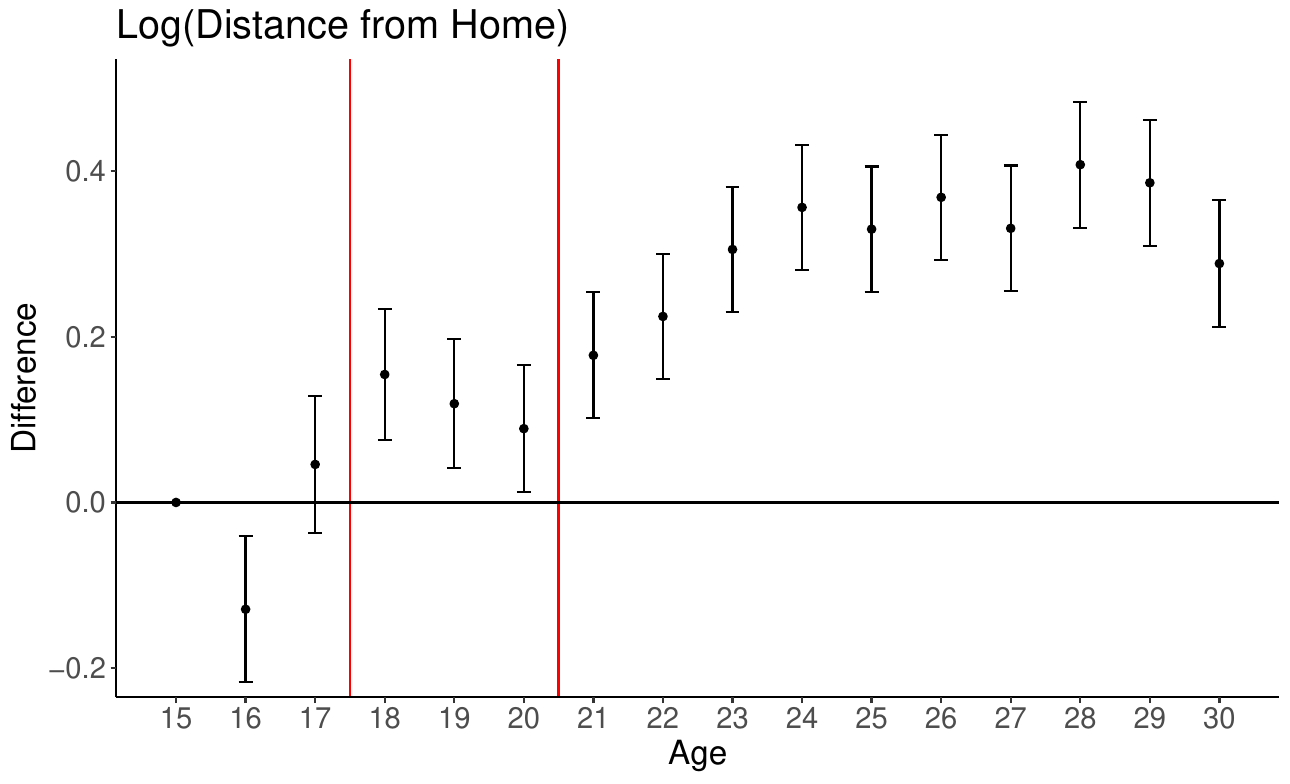}
    \begin{notes}
    This figure reports estimates of $\beta_a$, the effect of college-attendance on four individual outcomes (y-axis) by age (x-axis), as in the difference-in-differences specification in equation \ref{eq:did_ed}. Black intervals denote 95\% confidence intervals of the estimate, clustered at the individual level. Dashed vertical lines denote typical ages of college entry and exit. Estimates are pooled across cohorts (age of college entry) and come from a 1\% sample users in our Facebook sample.
    \end{notes}
\end{figure}

Figure \ref{fig: dd_col} reports estimates of $\beta_a$ across four individual outcomes. We find that by age 30 college attendance increases total network size, the number of connected cities, and the average phone price of friends by 20\%, 15\%, and 6\%, respectively. We also find that by age 30 the distance to home increases by 25\%, consistent with both human and social capital gains driving out-migration. Education may thus have additional returns in the labor market through its impact on network expansion and the resulting reduction in moving costs.

\section{Discussion}
\label{sec:discussion}
In this paper, we use aggregated and de-identified data from Facebook to show that social networks are an important dimension of internal migration. We find that a substantial portion of the gravity relationship in India can be explained by networks. A simple spatial equilibrium framework suggests expanding or reallocating social networks across space may deliver significant economic gains, especially so for low-income populations. Our data and surveys suggest both economic and emotional support are important mechanisms for this reliance on social networks. Finally, we find that college attendance leads to significant gains in social capital.

Our conclusions are limited in several ways. We use aggregated and de-identified microdata from Facebook that allows us to observe both migration and social networks with fine granularity, at the individual level. To our knowledge, this is the most comprehensive and detailed data on either individual migration patterns or social networks in India. However, our sample differs from India's broader population, and is generally located in higher income regions and has a higher propensity to migrate. In addition, we use a restricted sample of Facebook users for which we observe consistent migration data over four years. In that sense, parameter estimates are conditional on this selected sample. When re-weighting our data to match India's income distribution across regions, we find that resulting estimates are comparable. However, there remains unobserved differences even within region and income groups between our sample and that of India's population, which we cannot adjust for and may further impact estimates.

In addition to these data considerations, our model is limited in two key ways. First, we do not account for selection on skill in our model of migration. In particular, individuals choose locations based on average wages and not expected wages conditional on their skill. In general, migrants come from the right tail of the income and education distribution of their hometown, and thus may face expected wages that differ from average wages. Indeed, \cite{bm19} find that selection explains a substantial portion of the observed wage heterogeneity in Indonesia and therefore reduces the potential gains from lowering moving costs. In this paper, we thus interpret our wage elasticity as capturing labor supply responses to (and resulting estimates of WTP as relative to) average wages and not individual wages. In equilibrium, we anticipate selection forces would decrease the potential gains of social network expansion.

Second, we do not account for dynamics or endogenous network formation in our spatial equilibrium framework. Under various counterfactuals in period $t$, we solve for a new equilibrium by finding the distribution of wages and population that satisfy equilibrium conditions in period $t+1$. We do not attempt to model how and at what rate that economy moves to the new equilibrium. In particular, as individuals change location decisions in period $t+1$, this impacts the distribution of networks in $t+1$, which may cause further migration in period $t+2$. In addition, social interactions between individuals may expand social networks, particularly so in areas with more migration and high population density. As \cite{munshi20} points out, dynamic models of migration with social networks suggest substantially larger network effects due to dynamic complementarities. We anticipate these dynamic and network formation forces would increase the potential gains of social network expansion. Incorporating both selection and dynamics into this framework would therefore be a fruitful avenue for further research.

\newpage
\printbibliography
\end{document}